\documentclass[12pt,epsf]{article}

\usepackage{amsmath,amssymb,graphicx,cite}

\newcommand{\beq}{\begin{eqnarray}}
\newcommand{\eeq}{\end{eqnarray}}

\title{
\vspace*{5mm} {\huge \bf SUSY Hidden in the Continuum}
\vspace*{0.5cm}
\author{\textbf{Haiying Cai, Hsin-Chia Cheng, Anibal D.~Medina}\\
 \textbf{and John Terning}\\
\\
\normalsize\emph{Department of Physics, University of California,
One Shields Ave. Davis, CA 95616, USA}\\}}

\date{\today}
\begin{document}
\setcounter{page}{0} \maketitle

\vspace*{0.5cm} \maketitle 
\begin{abstract}
  We study models where the superpartners of the ordinary particles have
 continuous spectra rather than being discrete states, which can occur
 when the supersymmetric standard model is coupled to an approximately
 conformal sector.  We show that when superpartners that are well into the
 continuum are produced at a collider they tend to have long decay chains that
 step their way down through the continuum, emitting many fairly soft
 standard model particles along the way, with a roughly spherical energy distribution in
 the center of mass frame.
\end{abstract}

\thispagestyle{empty}

\newpage

\setcounter{page}{1}

\section{Introduction}

In this paper we take the first steps in studying the collider
phenomenology of supersymmetric models that have some approximate
conformal symmetry, which is  reflected in the fact that the
superpartners of the standard model (SM) particles have a
continuous spectrum~\cite{Continuum} rather than being discrete
states. The superpartners can have continuous spectra when the
supersymmetric standard model is coupled to a conformal field
theory (CFT) sector~\cite{Continuum} which gives rise to so-called
``unparticle'' behavior.\footnote{Unparticles are fields with a
continuous spectrum \cite{Georgi,Georgi2}, possibly with a with a
mass gap \cite{Fox,coloredunparticles}, whose two point functions
exhibit a nontrivial scaling behavior.} A mass gap is generated
for the superpartners of (and also the excitations of) the SM
particles if the conformal symmetry is softly broken in the
infrared (IR). This scenario is most easily modeled in a
five-dimensional (5D) anti-de Sitter (AdS) space using the AdS/CFT
correspondence~\cite{AdSCFT,Cacciapaglia}. The 5D AdS space is cut
off by an ultraviolet (UV) brane where supersymmetry (SUSY) is
broken. In the absence of an IR brane, the SM fields propagating
in the bulk will have continuous spectra. Such a theory would have
been ruled out if the spectra continue down to zero mass. However,
a mass gap for nonzero-modes can be generated by introducing a
soft wall in the IR, which can be parameterized by a position
dependent bulk mass term or a dilaton field with an appropriate
profile.  The boundary conditions on the UV brane remove half of
the fermion zero-modes so that the SM chiral fermions can be
obtained. In the supersymmetric limit, the 4D  theory consists of
SM particles and their superpartners as the zero-modes, plus a
continuum of Kaluza-Klein (KK) excitations starting from some gap
for each field. After SUSY breaking is introduced on the UV brane,
the zero modes of the superpartners are lifted while the mass gaps
of the continuum excitations are not affected since they are
determined in the IR. Depending on the parameters, for large
enough SUSY breaking the zero mode of the superpartner can merge
into the continuum and only a continuous spectrum for the
superpartner is left.

As one can imagine, the collider phenomenology could
be quite complicated with continuous spectra. Calculations of production
cross-sections have already been discussed in \cite{coloredunparticles}, so here
we will focus on decay processes. Imagine that
some highly excited mode in the gluino continuum is produced at
the Large Hadron Collider (LHC), it can decay to a squark with an arbitrary mass in
the continuum between the squark gap and the initial gluino mass (neglecting the mass of the
emitted quark). The squark can then decay back to a
gluino as long as the squark mass is above the gluino gap.
An obvious question is: does a continuum decay prefer to occur for small
mass differences or large mass differences? If the gluino prefers
to decay to a squark with an invariant mass close to its own, then the jet
emitted in the decay will carry a relatively small amount of energy, and
there will be many steps of decays before it reaches the bottom of
the spectrum. The events will contain a high multiplicity of soft
visible particles, which can be quite challenging at hadron colliders.
On the other hand, if the gluino prefers to decay to the squark
near the bottom of the spectrum, then we expect only a few decay steps
and hard jets from the decays. The signal events in this case are
more like traditional SUSY models. Since the theory becomes conformal at
high energies, we expect that in the high invariant mass limit we will be closer
to the first picture \cite{Strassler,Polchinski,HofmanMaldacena,Csaki}. Here we would like to
make some more quantitative statements.

Explicit calculations for continuum superpartner events, however, are
not straightforward. Because the initial and/or final states are
not particles, the usual Feynman rules for particles are not
directly applicable. One way to avoid this problem is to introduce
a regularizing IR brane which makes the extra dimension compact,
then the continuum becomes discrete KK modes \cite{Stephanov} and we can perform
calculations just as in the particle case. The continuum limit is
obtained by taking the position of the IR brane  to
infinity. The physical results should not depend on this position, as
long as it is much larger than the length scale of the inverse mass gap so that the KK modes are dense enough to approximate
a continuum. Even in this case, one may worry about whether the
usual narrow-width approximation of splitting a decay chain into
steps with independent decays is a good one, as there are an infinite number of KK modes which
can make virtual (``off-shell") contributions, reflecting the fact that the continuum state do not have a mass shell to be on.
We will discuss the validity of this approach and compare full calculations with the narrow-width approximation.

This paper is organized as follows. We first review the 5D construction
of continuum superpartners, then plough into the details of the decay chains, showing that the narrow-width
approximation is generally valid in perturbative theories.  We then discuss the phenomenology and the parametric dependence of the
observable quantities and give conclusions.  A detailed check of the IR regulator is included in the Appendix.

\section{A Review of Continuum Superpartners}

Let us recall the setup used in Ref.~\cite{Continuum}, which will be
the starting point of our collider studies. We consider a 4D
SUSY theory with approximate conformal symmetry. Through the AdS/CFT
correspondence \cite{AdSCFT}
this can be modeled by a 5D AdS space. We take the metric
of the AdS$_5$ space to be
\beq d s^2 = \left( \frac{R}{z} \right)^2 \left( \eta_{\mu \nu} d
x^\mu d x^\nu - d z^2 \right)\,.\label{metric} \eeq
The space is cutoff at $z=z_{UV}=\epsilon$ by a UV
brane where SUSY is broken.
As is well known, such a theory can be described in the language of 4D $\mathcal{N}=2$
superfields, which implies that for each matter field its 5D $\mathcal{N} =1$
hypermultiplet $\Psi$ can be decomposed into two 4D $\mathcal{N} =1$ chiral
superfields $\Phi=\{\phi,\chi,F\}$ and $\Phi_c=\{\phi_c,\psi,F_c \}$,
with the fermionic Weyl components forming a Dirac fermion \cite{MartiPomarol}.
In the case of gauge fields, a 5D $\mathcal{N} =1$ vector superfield can be decomposed
into an $\mathcal{N} =1$ 4D vector superfield $V=\{A_{\mu},\lambda_1,D\}$ and
a 4D $\mathcal{N} =1$ chiral superfield $\sigma=\{(\Sigma+iA_5)/\sqrt{2},\lambda_2,F_{\sigma}\}$.

\begin{figure}[htb]
\centering
\begin{tabular}{lcr}
\includegraphics[scale=0.5]{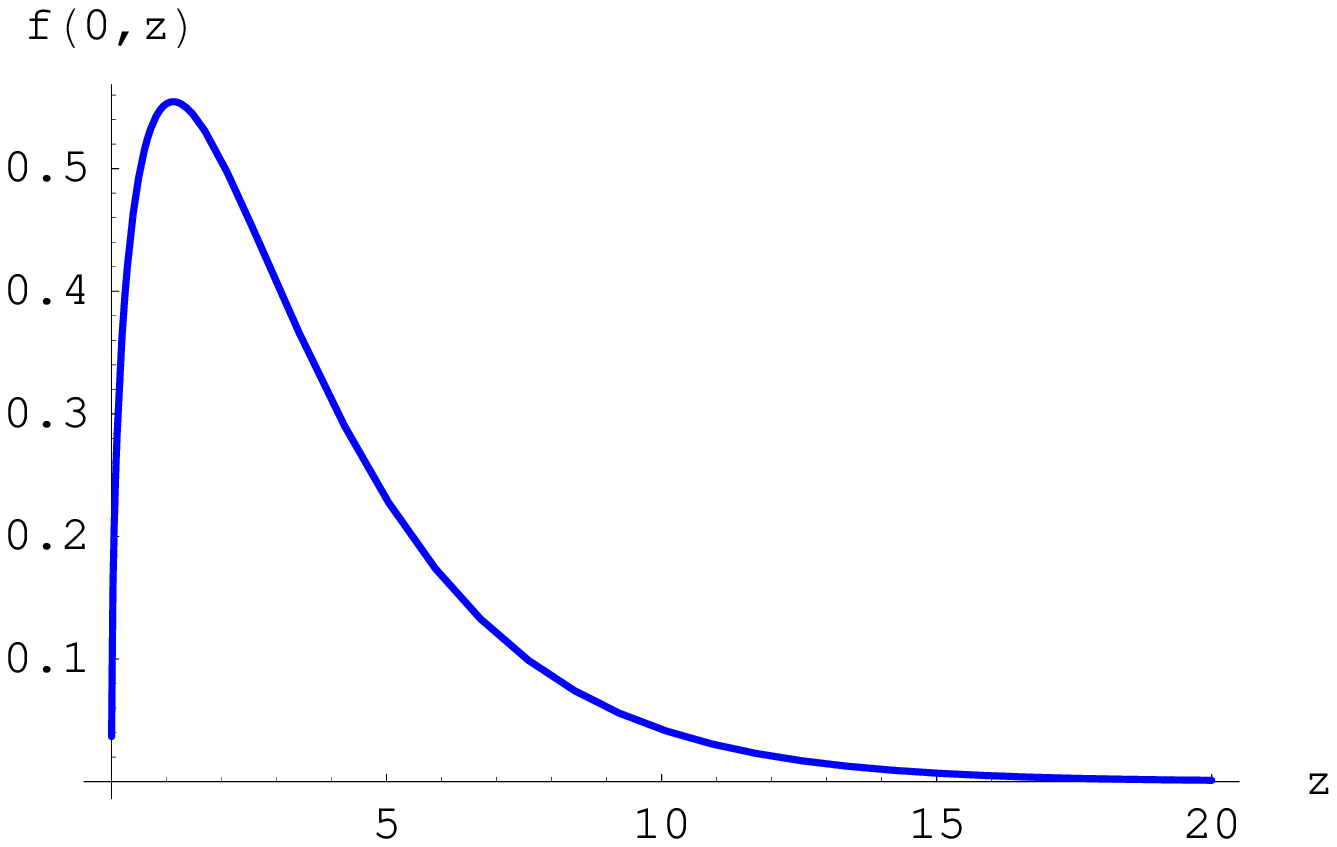}
&\qquad \qquad &
\includegraphics[scale=0.5]{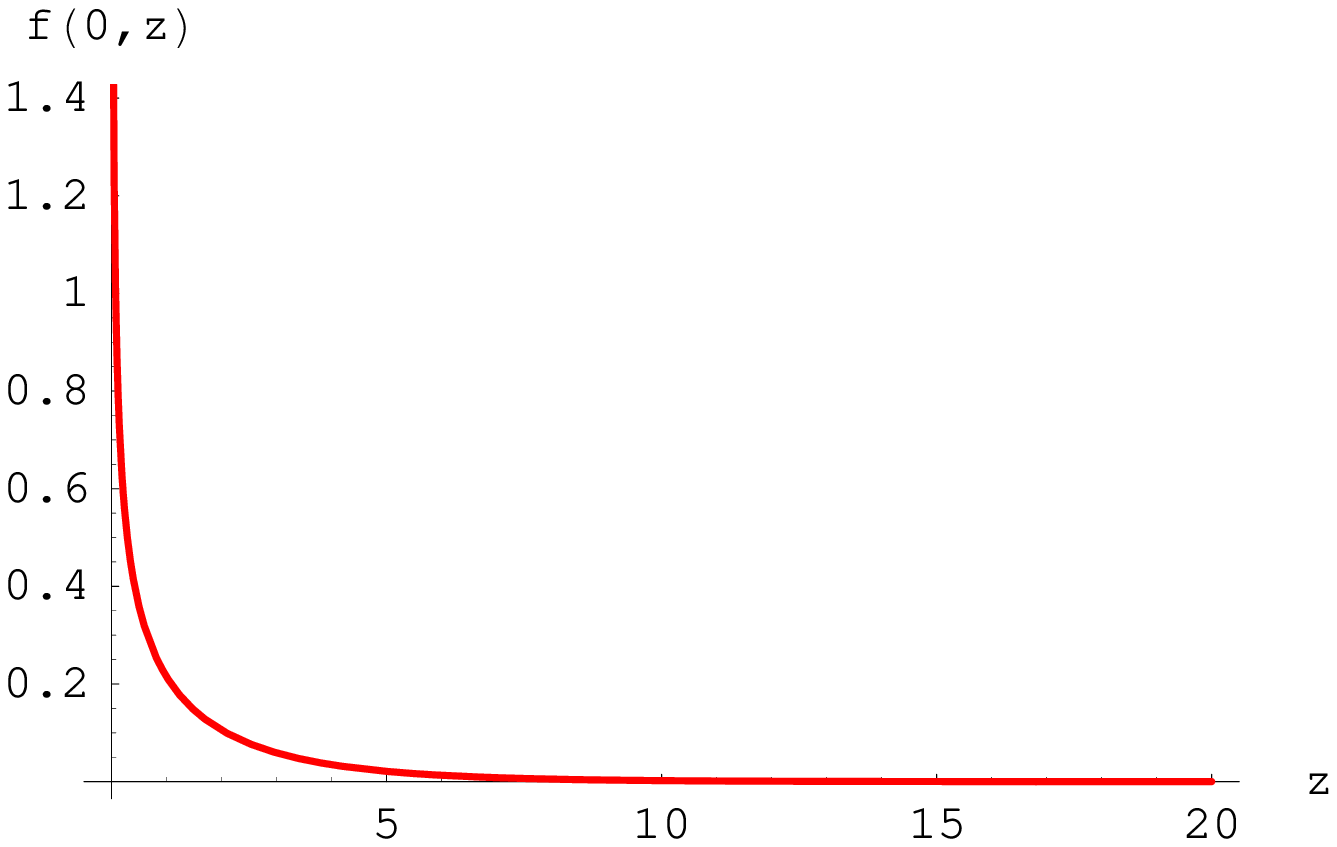}
\end{tabular}
\caption {Normalized zero mode profiles. In the left panel we take
 $c= - 0.45$, $ \mu_{f} = 0.4 ~\mbox{TeV}$ while in the right panel, we take
 $c = 0.45$, and $\mu_{f} = 0.4 ~\mbox{TeV}$.}\label{zeromode}
\end{figure}

As in the usual case of extra-dimensional theories, one can decompose the matter fields as:
\begin{eqnarray}
 \chi (p,z) & = &  \chi _{\rm{4}} (p)\left(\frac{z}{z_{UV}}\right)^2 f_L(p,z),  \qquad
 \phi (p,z) =  \phi _{\rm{4}} (p)\left(\frac{z}{z_{UV}}\right)^{3/2} f_L(p,z), \label{decom1}\\
 \psi (p,z) & = &  \psi _{\rm{4}} (p)\left(\frac{z}{z_{UV}}\right)^2 f_R(p,z), \qquad
\phi _c (p,z) =  \phi _{{\rm{c4}}}
(p)\left(\frac{z}{z_{UV}}\right)^{3/2} f_R(p,z)\label{decom2},
 \end{eqnarray}
where the relationships between scalar and fermion profiles are
provided by SUSY.
With a fermion bulk mass $m(z)= c+\mu_{f} z$, the bulk equations of motion give
\begin{eqnarray}
\frac{\partial^2}{\partial z^2}f_R+\left(p^2-\mu_f^2-2\frac{\mu_f
c}{z}-\frac{c(c-1)}{z^2}\right)f_R&=&0\label{2ndorder1},\\
\frac{\partial^2}{\partial z^2}f_L+\left(p^2-\mu_f^2-2\frac{\mu_f
c}{z}-\frac{c(c+1)}{z^2}\right)f_L&=&0 \label{2ndorder2}.
\end{eqnarray}
The solutions for $f_L(p,z)$ and $f_R(p,z)$
can then be expressed as linear combinations of the first order
and second order Whittaker functions:
\begin{eqnarray}
f_L(p,z) &=& a \,M( \kappa,\frac{1}{2} + c,2\sqrt {\mu_{f}^2  - p^2 } z)
+ b \, W( \kappa,\frac{1}{2} + c,2\sqrt {\mu_{f}^2  - p^2 } z)~, \label{eq:fl}
\\
f_R(p,z) &=& -a \, \frac{2(1 + 2c)\sqrt {\mu_{f}^2  - p^2 }}{p} M(
\kappa, - \frac{1}{2} + c,2\sqrt {\mu_{f}^2 - p^2 } z) \nonumber \\
&&-\, b \, \frac{p}{(\mu_{f} +\sqrt {\mu_{f}^2 - p^2 } )} W( \kappa, -
\frac{1}{2} + c,2\sqrt {\mu_{f}^2 - p^2 }z)~, \label{eq:fr}\\
\kappa&\equiv&- \frac{{c\,\mu_{f}}}{{\sqrt {\mu_{f}^2  - p^2 } }}~,
\end{eqnarray}
where $a$ and $b$ are coefficients which may depend on
four-momenta $p$ and that are determined by boundary and
normalization conditions. We will take all SM fields as
left-handed, so we demand that the right-handed chiral superfield
$\Phi_c(p, z)$ satisfies the Dirichlet boundary condition at $z =
z_{UV}$, {\it i.e.}, $\Phi_c(p, z_{UV}) = 0 $. The left-handed
chiral superfield $\Phi$, through the equations of motion,
satisfies modified Neumann boundary conditions.  As a result, only
the left-handed chiral  superfield has a normalizable zero-mode.
As shown in Ref.~\cite{Continuum}, the zero mode profile for the
left-handed field is given by
\begin{equation}
f^0_{L}(p,z)=\mathcal{N} (\mu_{f}, 0)z^{-c}e^{-\mu_{f} z},
\end{equation}
where the normalization factor $\mathcal{N} (\mu_{f}, 0)$ is obtained from
$~ \int_0^\infty { {f(0,z)} {f(0,z)^{*}} }
dz = 1$,
\begin{equation}
 \mathcal{N}(\mu_{f}, 0) = (2^{ - 1 + 2c} \mu_{f} ^{ - 1 + 2c} ~ \Gamma
(1 - 2c))^{-1/2}~. \label{enorm}
\end{equation}
In Fig.~\ref{zeromode}, we show the zero mode profiles for two values
of $c$. As can be seen, when $c$ is positive,
the zero mode is localized near the UV brane, while for $c<0$, the zero
mode is repelled away from the UV brane.

Similarly for gauge fields \cite{Continuum} we write:
\begin{eqnarray}
 \lambda_1 (p,z) & = &  \chi _{\rm{4}} (p) e^{uz} \left(\frac{z}{z_{UV}}\right)^2 h_L , \quad
 A_\mu  (p,z) = A_{\mu 4 } (p) e^{uz}  \left( \frac{z}{z_{UV}} \right)^{1/2} h_L ,\\
 \lambda_2 (p,z) & = &  \psi _{\rm{4}} (p)e^{uz} \left(\frac{z}{z_{UV}}\right)^2 h_R, \quad
\Sigma =  \phi _4 (p)e^{uz} \left( \frac{z}{z_{UV}} \right)^{3/2}
h_R\, ,
\end{eqnarray}
where $h_{L,R}$ represents $f_{L,R}$ evaluated at $c=1/2$ and $\mu_{g}$ is related
to the dilaton vacuum expectation value \cite{Cacciapaglia}, $\langle\Phi\rangle=e^{- 2 \mu_{g} z} / g_5$.

After including SUSY breaking on the UV brane, the zero mode of
the superpartner will be lifted,  and the SUSY breaking mass that
it acquires depends on the overlap of its wave function with the
UV brane. For $c$  close to +1/2, the zero mode of the
superpartner will acquire a SUSY breaking mass  near the full
strength ({\it i.e.}, comparable to the SUSY breaking on the UV
brane). On the other hand, if $c<0$, the SUSY-breaking mass of the
zero mode superpartner is suppressed relative to the SUSY breaking
on the UV brane. Because $c=1/2$ for the gauge field, we will take
all SM fields to have $c$ close to 1/2 in this paper in order to
have similar SUSY-breaking masses.

The spectrum of nonzero-modes is continuous without an IR cutoff.
As mentioned in the Introduction, it is convenient to introduce a
regulating IR brane at a large distance $z=z_{IR}=L$ so that we
can deal with discrete normalizable KK states \cite{Stephanov}. The continuum limit
is obtained by taking $L\to+\infty$.
The coefficients $a$ and $b$ of the wave functions in
Eq.~(\ref{eq:fl}-\ref{eq:fr}) can be obtained by imposing the
boundary condition on the IR brane and the normalization
condition,
\beq
 \tilde f_L (p,z) &=& \mathcal{N}_L (\mu_{f} ,p)\left ( M( \kappa ,\frac{1}{2} + c,2\sqrt {\mu_{f} ^2  - p^2 }
 z)\right. \nonumber \\ && \left. + ~ b \cdot W(\kappa,\frac{1}{2} + c,2\sqrt {\mu_{f} ^2  - p^2 } z) \right), \label{fL} \\
\nonumber \\
 \tilde f_R (p,z) &=& \mathcal{N}_R (\mu_{f} ,p)\left(\frac{{2(1 + 2c)\sqrt {\mu_{f} ^2  - p^2 } }}{{\mu_{f} - \sqrt
{\mu_{f} ^2  - p^2 } }} \cdot M( \kappa , - \frac{1}{2} + c,2\sqrt {\mu_{f}^2 - p^2 } z)\right. \\
&& \left. + ~ b \cdot W(\kappa, -\frac{1}{2} + c,2\sqrt {\mu_{f}^2 - p^2 }z) \right)\; . \label{fR}
\eeq
Imposing the boundary condition $f_R (p,z_{UV})=0$ we determine that,
\beq
b &=&  - \frac{{M( \kappa, - \frac{1}{2} + c,2\sqrt {\mu_{f} ^2  - p^2
} z_{UV})}}
 {{W( \kappa, - \frac{1}{2} + c,2\sqrt {\mu_{f} ^2  - p^2 } z_{UV})}}
 \cdot \frac{{2(1 + 2c)\sqrt {\mu_{f} ^2  - p^2 } }}{{\mu_{f} - \sqrt
{\mu_{f} ^2  - p^2 } }}\label{bcondition}\; ,
\eeq
and $ \kappa \equiv-c\mu_{f}/\sqrt {\mu_{f}^2  - p^2 } $.

We are especially interested in the behavior of the superpartners in the conformal limit at high energies.
In the limit of $p \gg \mu_{f} $, the normalization factor can be
expressed as
\beq \mathcal{N}_L  = -\mathcal{N}_R  = \left( {\frac{{2^{3 + 4c}
\pi \sec^2 (c\pi) }}{{\Gamma ( - \frac{1}{2} - c)^2}}} \cdot
z_{IR} \right)^{ - 1/2} . \label{NL}  \eeq
The gluino profiles can be found similarly by making the replacements
$c\to 1/2$ and $\mu_{f}\to \mu_{g}$ in Eqs.~(\ref{fL}--\ref{bcondition}).

\begin{figure}[t]
\centering
\begin{tabular}{cc}
\includegraphics[scale=0.53]{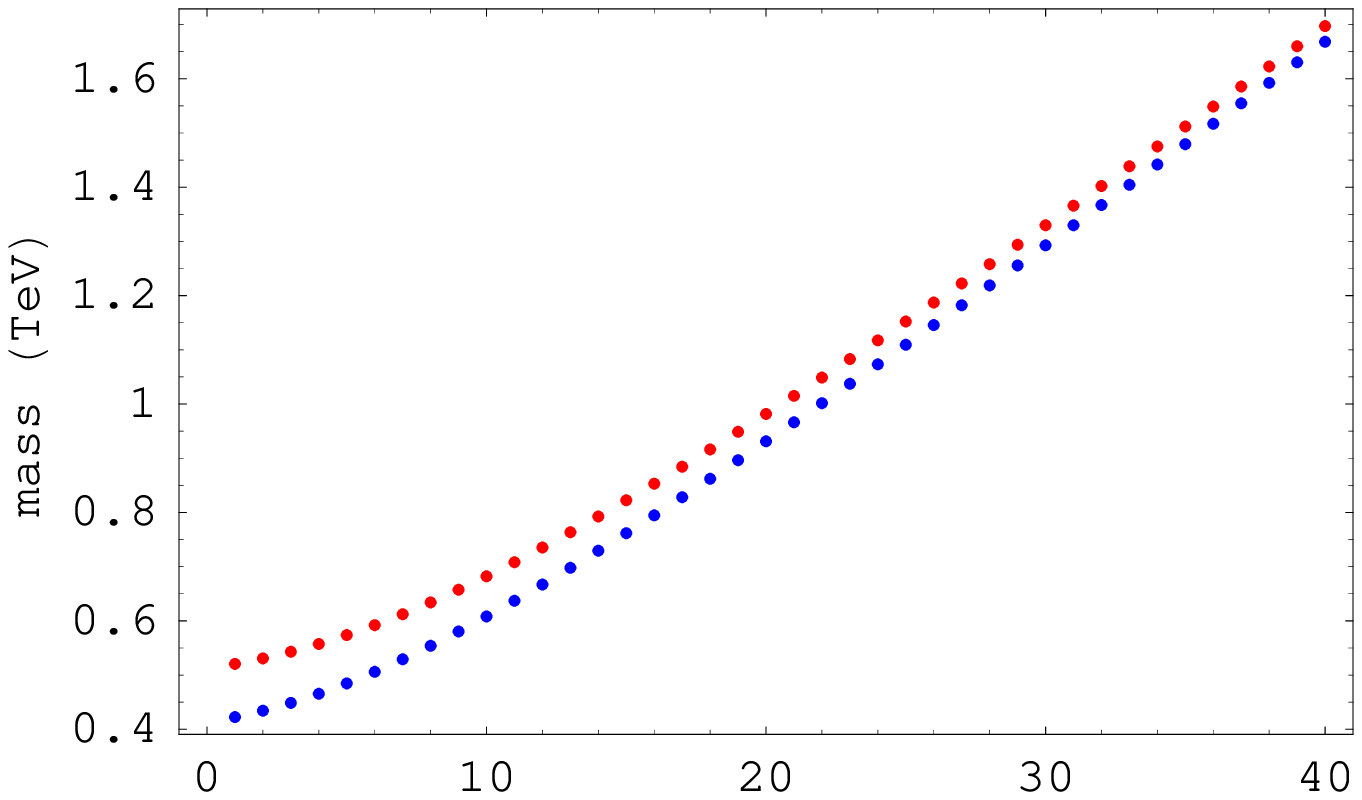}
&
\includegraphics[scale=0.53]{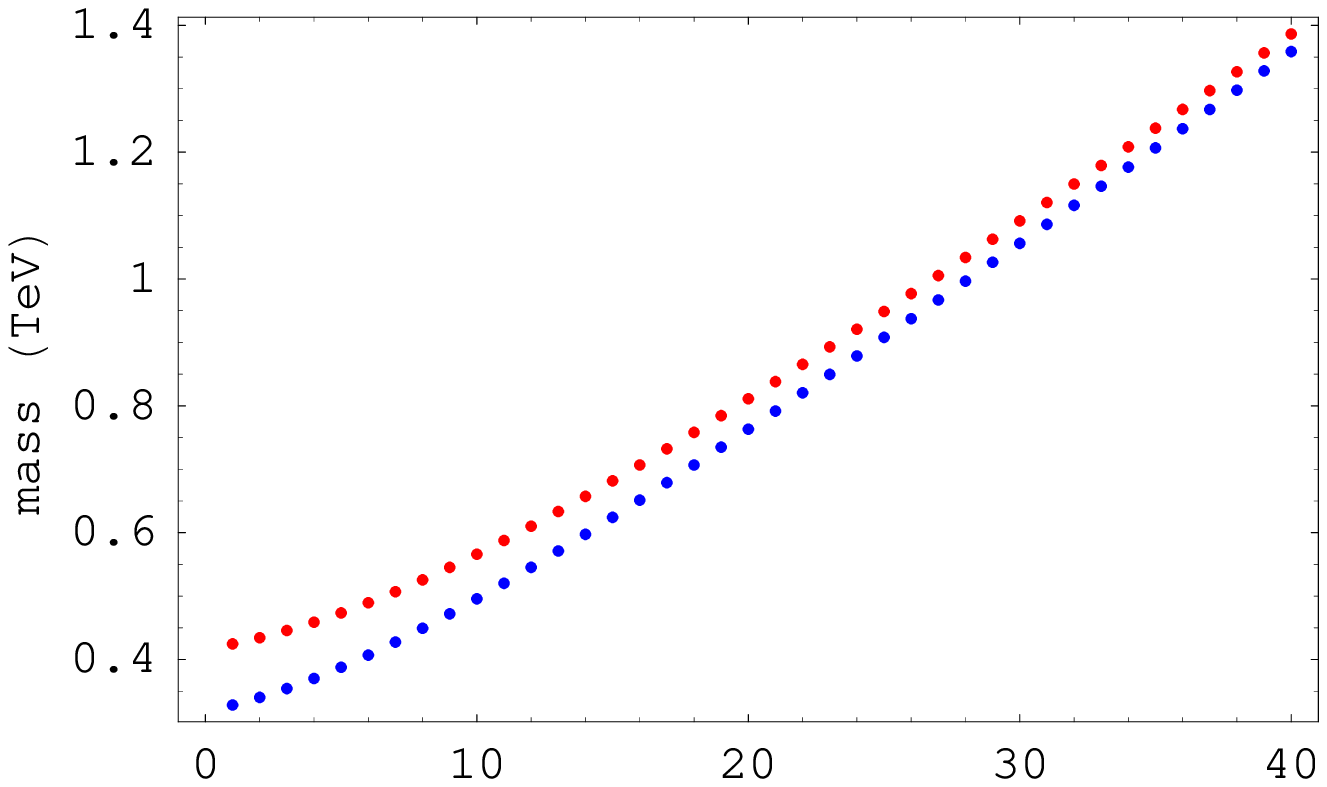}
\end{tabular}
\caption{\label{fig:kkspectrum} Spectra for the gluino  (blue
dots) and the squark (red dots). The plot shows mass versus KK mode number. The parameters are chosen to be
$z_{UV} = 10^{-3} ~ \mbox{TeV}^{-1}$, $z_{IR} = 80 ~
\mbox{TeV}^{-1}$, $c=0.5$, $\mu_{g} = 0.4 ~\mbox{TeV}$ and $\mu_{f} = 0.5
~\mbox{TeV}$ for the left panel, and $z_{UV} = 10^{-3} ~
\mbox{TeV}^{-1}$, $z_{IR} = 100 ~ \mbox{TeV}^{-1}$, $c = 0.5$, $\mu_{g}
= 0.3 ~ \mbox{TeV}$ and $\mu_{f} = 0.4 ~\mbox{TeV}$ for the right
panel.}
\end{figure}

The KK spectrum is determined by the boundary condition at the IR brane, $f_{R}(p,z_{IR})=0$,
which leads to:
\beq \frac{{M(- \frac{{\mu_{g}}}{{2 \sqrt {\mu_{g}^2 - p^2 } }} ,0 ,2\sqrt {\mu_{g}
^2  - p^2 }~ z_{UV})}}{{W(- \frac{{\mu_{g}}}{{2 \sqrt {\mu_{g}^2 - p^2 } }} ,0
,2\sqrt {\mu_{g} ^2  - p^2 }~ z_{UV})}}= \frac{{M(- \frac{{\mu_{g}}}{{2 \sqrt
{\mu_{g}^2 - p^2 } }} , 0 ,2\sqrt {\mu_{g} ^2  - p^2 }~ z_{IR})}}{{W(-
\frac{{\mu_{g}}}{{2 \sqrt {\mu_{g}^2 - p^2 } }} , 0 ,2\sqrt {\mu_{g} ^2  - p^2 }~
z_{IR})}} . \label{mass} \\ \nonumber \eeq
In the limits of interest, $\mu_{f},\mu_{g} \ll p \ll 1/z_{UV}$,
the left-hand-side  of Eq.(\ref{mass}) tends to zero, which
implies that the numerator of the right-hand-side of Eq.(\ref{mass}) satisfies:
\beq M(- \frac{{\mu_{g}}}{{2 \sqrt {\mu_{g}^2 - p^2 } }} , 0 ,2\sqrt {\mu_{g} ^2  -
p^2 }~ z_{IR}) \propto \cos (\frac{1}{4}\pi  - \sqrt {p^2  - \mu_{g} ^2
} z_{IR} ) = 0 . \label{spectrum}  \eeq
Solving Eq.~(\ref{spectrum}), we obtain an approximate expression for the KK masses,
\beq && m_n^2  \approx \mu_{g}^2  + (\frac{1}{4} + n)^2 \pi ^2 /z_{IR}^2  \qquad \mbox{with} \qquad  n =0, 1, 2, \cdots   .\eeq
Some sample spectra are shown in
Fig.~\ref{fig:kkspectrum}, where we have taken $z_{UV} = R = 10^{-3} ~
\mbox{TeV}^{-1}$ and $ c = 1/2$.  With mass gaps on the order of half a TeV, a 20 GeV mode spacing
is quite a good approximation to a continuum.

\section{Continuum Decay Chains and Narrow-Width Approximations}

Now we would like to study the decay of a continuum superpartner. The question is:
what are the characteristic features of the decay process? Does it
prefer to decay through multiple steps and emit soft particles
during the decays, or to decay directly down to the bottom of the
 spectrum together with a hard particle? What is the
typical energy of the visible particles produced in the decay
chain relative to the energy and other parameters of the superpartner? As mentioned earlier, in order to use the calculation
techniques developed for particles, it is convenient to include a
regularizing IR brane to discretize the continuum into KK modes \cite{Stephanov},
so that we can deal with ordinary particles in the initial and
final states. To study the validity of the narrow-width approximation of the decay chain, which splits
the process into a sequence of  independent decays at each step,
we compare the result from a calculation of
a sequence of real two 2-body decays with that of the 3-body decay which includes
all the contributions from the virtual intermediate continuum superpartner.
We assume that we start with an initial state with an
invariant mass much higher than the mass gaps and the
SUSY breaking mass. To simplify the calculations, we will ignore
the SUSY-breaking and the masses of the SM particles. We find
that, in the case that the calculations are reliable ({\it i.e.},
independent of $z_{IR}=L$ as long as $L \gg p^{-1}, \mu_{f}^{-1},
\mu_{g}^{-1}$), the virtual contributions are indeed smaller than the
real contributions so that the narrow-width approximation is reliable.

\subsection{ The 2-body gluino decay calculation}
\begin{figure}[t]
\centering \vspace{15pt}
\includegraphics[scale=1.0]{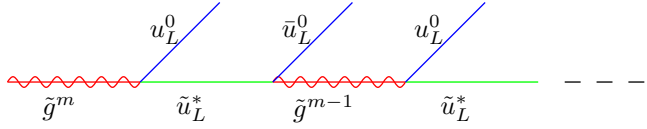}
\vspace{5pt}

\caption {Feynman diagram for sequential KK gluino decay.}
\label{feyn}
\end{figure}
Once we include a regulating IR brane to make the spectra
discrete, the calculation of a 2-body decay should be
straightforward. We consider an initial state of a gluino of KK
level $m$, decaying in its rest frame to a quark and a KK
squark of level $n$, $\tilde{g}^m\to u^0_L\tilde{u}^{*,n}_L$. Since we ignore
the quark mass, the decay can occur to any KK squark lighter
than the initial gluino.  The energy of the emitted quark
depends on which level of the KK squark the gluino decays to. To
calculate the decay rate, we need the coupling between the
gluino, squark, and the quark. The effective interaction of
the KK gluino, KK squark and the quark in momentum space is
given by:
\beq S_{eff}  = c(p_m, q_n) \int {\frac{{d^4 p_n }}{{(2\pi )^4 }}}
\frac{{d^4 q_m }}{{(2\pi )^4 }}~ u^0_L(p_m-q_n)
~\tilde{u}^{m,*}_L(q_n )~\tilde{g}^n (p_m) ~, \eeq
where $c(p_m, q_n)$ is the vertex coefficient which can be
computed by integration of the gluino, squark and quark 5D
profiles over the fifth dimension.  The expression is:
\beq c (p_m, q_n) &=&  \mathcal{N}_{\tilde g} (\mu_{g}, p_m) ~
\mathcal{N}_{\tilde u }(\mu_{f}, q_n)^{*} ~ \mathcal{N}_{u}(\mu_{f}, 0) ~
g_5\int_{z_{UV} }^{z_{IR}} {dz} \left( {\frac{z_{UV}}{z}}
\right)^5
 ~ \left(
{\frac{z}{z_{UV}}} \right)^2 e^{ - \mu_{f} z} z^{ - c}\nonumber \\
\nonumber \\ &&  \left( {\frac{z}{z_{UV}}} \right)^{3/2}
{f}_L(q_n,z)^{*}
  ~ e^{uz} \left( {\frac{z}{z_{UV}}} \right)^2
{h}_L(p_m,z)\; , \label{coupling}\eeq
where $g_5$ is the 5D gauge coupling, which only enters the
calculation as an overall factor. The 4D gauge coupling $g_4$ is
related to the 5D gauge coupling $g_5$ by integrating over the zero
mode profiles in the kinetic term for the gauge fields coupled with
the dilaton. In that case one finds \cite{Cacciapaglia} that,
\beq g_4^2  = \frac{g_5^2}{z_{UV}}  \frac{1}{{\log (1/(2 z_{UV}
\mu_{g})) - \gamma_E )}}, \eeq
where $\gamma_E$ is Euler's constant. Since we are interested
in the gluino decay into a quark and a squark, we consider
the QCD gauge coupling which equals
$ \alpha_s(m_{Z})= 0.1184 $ at the $Z$-mass. If we take $z_{UV} = 10^{-3} ~
\mbox{TeV}^{-1} $ and the nominal value for the gluino mass gap of $\mu_{g}=0.1$ TeV,\footnote{The dependence
on the gluino mass gap $\mu_{g}$ is only logarithmic, reflecting the gauge coupling's running.}  the 5D strong gauge coupling assumes
the dimensionful value $g_5 = 0.108~  \mbox{TeV}^{-1/2} $.

An immediate observation from Eq.~(\ref{coupling})  is that a
finite coupling in the limit $z_{IR}\to+\infty$ is obtained only
for $\mu_{f}>\mu_{g}$. For $\mu_{f} < \mu_{g}$, the integral is dominated by the large
$z$ region and it blows up as we take $z_{IR}\to+\infty$. In this
case we cannot perform a sensible calculation. One can
see that the exponentially growing factor comes from the dilaton
profile for the KK gluino, which is a result of that the zero
mode gauge field has to be a constant along the fifth dimension
due to gauge invariance. A similar situation happens in the
Randall-Sundrum 2 (RS2) scenario~\cite{RS2}. If one calculates the
self interactions among KK gravitons in RS2 with a regulating IR
brane, one also finds that the coupling blows up as one takes the
IR brane to infinity. In the 4D CFT picture, these KK
gravitons correspond to the conformal bound states. The large
coupling just means that these bound states are strongly coupled.
These divergent couplings can never make any physical process
involving UV zero-modes
infinite. As is well
known, the KK picture sometimes can give misleading results when
locality in the extra dimension is involved.
 The point is that any process starting with zero-modes
localized on the UV brane will mostly be sensitive to the physics
near the UV brane due to the locality in the extra dimension. The
``nonlocal'' process of producing KK gravitons in the deep IR
region must be suppressed due to interference of various diagrams
even though each of them can have a large coupling. In our theory,
all zero-modes are localized near the UV brane (for $c\sim 1/2$),
so all physical processes originated from the zero-modes should
also only be sensitive to the physics near the UV brane. The
divergent coupling in the $\mu_{f} < \mu_{g}$ case just means that
the na\"ive calculations done in the KK picture are not valid. On
the other hand, for $\mu_{f} >\mu_{g}$ the integral is dominated
by the region near the UV brane, and the calculations are
trustworthy. For the same reason, the coupling between a KK
gluino, a KK squark, and a KK quark is also dominated by the IR
region and diverges when the IR brane is moved to infinity.
However, since the wave functions of the KK quarks are suppressed
in the UV region, the decay to a KK squark plus a KK quark should
also be suppressed if the initial KK gluino was produced in the UV
region in the first place. To avoid such complications, we will
restrict our study to the $\mu_{f} > \mu_{g}$ case and consider
decays to the quark zero mode only in this paper.

After obtaining the coupling, it is straightforward to calculate
the decay rate to each KK squark. To compare with the continuum
limit, we express the result in terms of the differential decay
rate as a function of the energy of the outgoing quark:
\beq \frac{d\Gamma_{\tilde{g}^m \rightarrow u^0_L\tilde{u}^{n,
*}_L}}{d E_{u^0_{L}}}&\approx&  \sum_{n\in\;\mu_{f}<E_n<E_m}\frac{\Delta
\Gamma_{\tilde{g}^m \rightarrow u^0_L\tilde{u}^{n,*}_L}}{\Delta
E_{u^0_{L}} }\nonumber \\
&\approx& \sum_{n\in\;\mu_{f}<E_n<E_m}\frac{{c(p_m,q_n)c^\dag
(p_m,q_n)}}{{\Delta E_{u^0_{L}}}}\frac{E^2_{u^0_{L}}}{{4\pi p_m
}}\frac{{\rm Tr}[t^at^a]}{8} \label{twobody}\eeq
where $t^a$ are the SU(3) generators in the fundamental representation and $p_m \equiv \sqrt {p_m^2 } $. From conservation of
energy-momentum we have,
\begin{equation}
 E_{u^0_{L}}  = \frac{1}{2} p_m \left( {1 - \frac{{q_n^2
}}{{p_m^2 }}} \right),
\end{equation}
which implies
\begin{equation}
  \Delta E_{u^0_{L}}  =
\frac{1}{2} p_m \left(1 - \frac{{q_{n-1} ^2 }}{{p_m^2 }}\right) -
\frac{1}{2} p_m \left(1 - \frac{{q_{n} ^2 }}{{p_m^2 }}\right)\;.
\end{equation}
\begin{figure}[t]
\centering
\includegraphics[scale=0.6]{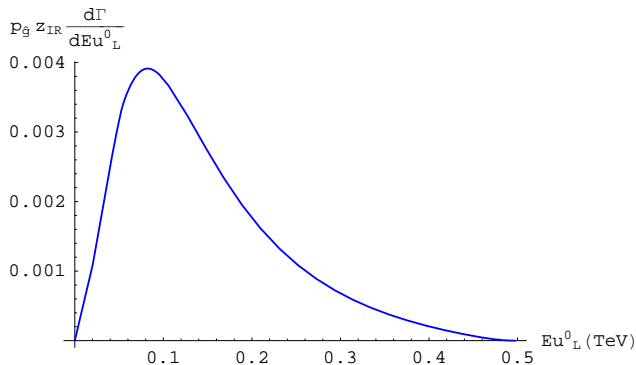}

\caption {Gluino differential decay rate with respect to the first
quark energy. The initial KK gluino mass is $ p_{\tilde g}=1.15
~\mbox{TeV}$ and we use $g_5 = 0.108~\mbox{TeV}^{-1/2}$.  We
choose $z_{UV} = 10^{-3} ~ \mbox{TeV}^{-1}$, $z_{IR} = 100 ~
\mbox{TeV}^{-1}$, $c=0.5$, $\mu_{g} = 0.3 ~\mbox{TeV}$ and
$\mu_{f} = 0.4 ~\mbox{TeV}$.}\label{diffrate}
\end{figure}
A typical differential decay rate as a function of the outgoing
quark energy is  shown in Fig.~\ref{diffrate}. We have normalized
the decay rate by a factor $p\cdot z_{IR}$ as we do with all
differential decay rate figures hereafter, in order to cancel the
unphysical IR dependence $\Delta m_n=\pi/z_{IR}$ in the continuum
limit.\footnote{It is easy to see that the coupling is
proportional to $L$ from the normalizations of the gluino and
squark wave functions. The differential decay rate of a single KK
gluino is proportional to $(1/L)^2 \times L=1/L$ as the density of
the final KK squarks is proportional to $L$. This is related to
the fact that the overlap of a single KK gluino wave function with
the UV region where the zero modes are located is also
proportional to $1/L$. In reality the initial KK gluino will be
produced with some energy range and the number of KK gluinos in
that energy range is again proportional to $L$, which cancels the
unphysical $1/L$ dependence in the differential decay rate of a
single KK mode.} At small quark energies, the decay rate is
suppressed by phase space, while at large quark energies, the
decay rate is suppressed by the couplings due to the small overlap
between the gluino and squark wave functions of large KK level
differences. Overall we see that the suppression due to the
coupling is stronger and the outgoing quark has a relatively soft
spectrum. The suppression of high energy emissions is exactly the
behavior that is expected in a CFT
\cite{Strassler,Polchinski,HofmanMaldacena,Csaki}. Starting with
very high KK modes this leads to approximately spherical events
\cite{Csaki}. A more detailed discussion of the quark spectrum and
its parameter dependence will be presented in the next section
after the narrow-width approximation  is justified.

\subsection{The 5D mixed position-momentum propagator }

{}From the previous subsection we see that the result of the
2-body calculation shows that  the gluino prefers to decay to the
squark with a mass not far below the gluino mass. This means that
the resulting squark is likely to decay again back to the gluino
and the whole process will involve a long decay chain. An
important question is whether the sequence of decays can be treated independently and
correlations implied by the full intermediate propagator
can be safely neglected. This question is less
trivial for the continuum case than the usual particle case
because there are an infinite number, a continuum, of intermediate
states which can give contributions at each step.  To study this problem, we
consider the 3-body gluino decay process $\tilde{g}^m\to u^0_L
u^{0*}_{L}\tilde{g}^n$ where the continuum squark is in the intermediate
state. We will calculate the contribution from the virtual (``off-shell")
intermediate squark while using real KK modes for the initial and final states.

In the KK picture with a regulating IR brane, the intermediate
squark propagator  is simply a sum of the particle propagators of
all KK levels \cite{Stephanov}. The propagator of an individual scalar particle is
\beq
\lim_{\epsilon\rightarrow 0^+} \frac{i}{q^2  - m_{n}^2  + i \epsilon} = \pi \delta (q^2  - m_{n}^2 ) +i\,{\mathcal P} \frac{1}{q^2 -
m_{n}^2}
\label{PPart}
 \eeq
where ${\mathcal P}$ denotes the Cauchy principal value. The delta function (the
real part) represents the phase space for a real  intermediate
particle (the ``on-shell" contribution), while the second term (the
imaginary part) represents
the contribution of a virtual intermediate state (the ``off-shell" contribution). However, the KK picture is not the most
convenient one to perform calculations with virtual intermediate states as it involves
an infinite sum of singular functions. We know that in the
continuum limit, the infinite sum of  KK propagators simply becomes
the unparticle propagator \cite{Stephanov}, and   the series of poles on the positive
real axis of $p^2$ merge into a branch cut of the unparticle
propagator. Therefore, we will employ the full unparticle
propagator for the intermediate state in our calculation. In this
subsection we derive the unparticle propagator in mixed
position-momentum space \cite{Falkowski}.

The left-handed scalar propagator satisfies
the equation of motion for the left-handed profile, with a delta-function source,
\beq \left( {-\partial _z^2  + \frac{3}{z}\partial _z  + (c^2  + c
- \frac{{15}}{4})\frac{1}{{z^2 }} + \frac{{2c\mu_{f} }}{z} + \mu_{f} ^2 -
p^2 } \right) i P(p,z,z') = \left( \frac{z}{z_{UV}} \right)^{ 3}
 \delta (z - z')  , \label{Green}\eeq
and the UV boundary condition for the propagator is
\beq \left. {\left( {\partial _z  + \frac{1}{z}( - \frac{3}{2} + c
+ \mu_{f} z)} \right)P (p,z,z')} \right|_{z = z_{UV}}  = 0  .\label{Greenbc} \eeq
To connect to the continuum limit, with outgoing boundary conditions, we demand that the propagator
is exponentially damped \cite{Falkowski} at large Euclidean momenta:
\begin{equation}
P(i p,z_{IR},z')|_{p\to+\infty,z'<z_{IR}}\to e^{-p\,z_{IR}} .
\end{equation}
In order to solve Eq.~(\ref{Green}--\ref{Greenbc}) for $P(p,z,z')$, we look for solutions in the
regions $z> z'$, $P_ >  (p,z,z')$,  and $ z < z'$, $P_ <  (p,z,z') $, and use matching boundary conditions at $ z= z'$,
\beq
 && \left. {P_ <  (p,z,z') - P_ >  (p,z,z')} \right|_{z = z'} = 0, \\
&&  \left. { {\partial _z P_ < (p,z,z') - \partial _z P_ >
(p,z,z')} } \right|_{z = z'} = -i \left( \frac{z}{z_{UV}}
\right)^{3} . \eeq
We can write the general solution to Eq.~(\ref{Green})
in terms of two independent solutions to the homogeneous equation \cite{Falkowski}, $K(p,z)$ and
$S(p,z)$:
\beq
 K(p,z) &=& \left( {\frac{z}{z_{UV}}} \right)^{3/2}
 \frac{{W(\kappa ,\frac{1}{2} + c,2\sqrt {\mu_{f} ^2  - p^2 } z)}}
 {{W(\kappa ,\frac{1}{2} + c,2\sqrt {\mu_{f} ^2  - p^2 } z_{UV})}}\;,  \\ \nonumber \\
 S(p,z) &=& \left( {\frac{z}{z_{UV}}} \right)^{3/2} \frac{1}{{2\sqrt {\mu_{f} ^2  - p^2 } }}
 \frac{{\Gamma (1 + c - \kappa )}}{{\Gamma (2 + 2c)}} \nonumber \\
&& \left( M(\kappa ,\frac{1}{2} + c,2\sqrt {\mu_{f} ^2  - p^2 } z) ~
W(\kappa ,\frac{1}{2} + c,2\sqrt {\mu_{f} ^2  - p^2 } z_{UV}) \right.
\nonumber
\\ && \left. - W(\kappa ,\frac{1}{2} + c,2\sqrt {\mu_{f} ^2  - p^2 } z) ~ M(\kappa
,\frac{1}{2} + c,2\sqrt {\mu_{f} ^2  - p^2 } z_{UV})  \right) \label{SN}, \eeq
which satisfy the following boundary conditions:
\beq K(p,z_{UV}) = 1,\qquad K(i p,z)|_{p\to+\infty}\to e^{-pz},  \qquad S(p,z_{UV})=0, \qquad
S'(p,z_{UV})=1. \eeq
In the region $z < z' (z> z')$, the general solution can be written as
\beq P_ {< (>)} (p,z,z') &=& a_ {<(>)} ~ K(p,z) + b_ {<(>)} ~ S(p,z) .\eeq
The boundary condition at the UV brane, Eq.~(\ref{Greenbc}), fixes
the ratio of $a_</b_<$  to be proportional to the kinetic function
$\Sigma_{F_c}(p)$ already encountered in Ref.~\cite{Continuum}:
\beq \frac{a_< }{ b_<}=\frac{\Sigma_{F_c}(p)}{z_{UV}}=
\frac{(\mu_{f} +\sqrt {\mu_{f}^2  - p^2 })}{p^2 } \frac{{W\left( {
- \frac{{c\mu_{f}}}{{\sqrt { - p^2  + \mu_{f}^2 } }},\frac{1}{2} +
c,2\sqrt { - p^2 + \mu_{f}^2 }~ z_{UV}} \right)}}{{W\left( { -
\frac{{c\mu_{f}}}{{\sqrt { - p^2  + \mu_{f}^2 } }},\frac{1}{2} -
c,2\sqrt { - p^2  + \mu_{f}^2 }~ z_{UV}} \right)}}~,\eeq \beq
\Sigma_{F_c}(p) &=&z_{UV} \cdot
\left(\frac{R}{z_{UV}}\right)^3\frac{1}{p}\frac{f_L}{f_R}~. \eeq
The kinetic function $\Sigma_{F_c}(p)$ not only fixes the spectral density, but it is
also essential in determining the phase space.
Using the required IR behavior of $P (p,z,z')$, we conclude that
$P_ >  (p,z,z')$ can only be proportional to $K(p,z)$, ({\it
i.e.}, $b_ > = 0 $). At this stage we use the matching conditions
at $z=z'$, so that in the range $ z < z'$, the left-handed squark
propagator can be expressed as \cite{Falkowski}:
\beq i\,P(p,z,z') &=& \frac{\Sigma_{F_c}(p)}{z_{UV}}K(p,z)K(p,z')
- S(p,z)K(p,z')~. \label{pro} \eeq
 For the case $z>z'$ we just interchange $z\leftrightarrow z'$ in Eq.~(\ref{pro}).

Obviously for $p^2 > \mu_{f}^2$,
$\sqrt{\mu_{f}^2- p^2}$ becomes imaginary, so the propagator has a
branch cut on the real axis for $p^2 >\mu_{f}^2$, and the
discontinuity is just twice the real part. This discontinuity corresponds
to a real intermediate unparticle (the ``on-shell'' contribution).  From the
analogy of the particle propagator (\ref{PPart}) we interpret the
imaginary part of the unparticle propagator as the virtual (``off-shell")
contribution to the 3-body decay process, while the real part (the discontinuity)
corresponds to the phase space of the unparticle. In fact, one can
use this phase space to calculate directly the real 2-body
differential decay rate considered in the previous subsection,
instead of summing over KK modes in a discretized theory. In the
Appendix we show the equivalence of the two pictures in the limit
that the IR regulator is removed,
$L\to \infty$. The numerical results for the 2-body differential
decay rates using the two approaches also agree well for large
$L$.

\subsection{The 3-body gluino decay}

With the unparticle propagator derived in the previous subsection,
we can compute the virtual contribution to the 3-body decay
process, $\tilde{g}^m\to u^0_L u^{0*}_{L}\tilde{g}^n$. We take
both the initial and final gluinos to be KK states
in an IR regularized theory, but use the unparticle propagator for
the intermediate state with only the imaginary part corresponding to the
virtual contribution. It is straightforward to find the
amplitude squared for the virtual 3-body process by integrating
the imaginary part of the propagator $P (q, z, z')$ derived in
Eq.(\ref{pro}) over the positions of the two vertices in the extra
dimension,
\beq \left| {M(\tilde{g}^m\to u^0_L u^{0*}_{L}\tilde{g}^n )}
\right|^2 = 4 g_5^4 ~ |v (p_{\tilde{g}^m}, q,
p_{\tilde{g}^{n}})|^2 ~ ( p_{\tilde{g}^{m}}.p_{u^{0*}_{L}}) ~ (p_{u^0_L}.p_{\tilde{g}^n} )~.
\label{amplitude} \eeq
We have labeled the 4-momenta of the initial, final, and
intermediate states as follows: $p_{\tilde{g}^{m}}$ for the
initial gluino $\tilde{g}^m$, $p_{\tilde{g}^{n}}$  for the
outgoing gluino $\tilde{g}^n $, $p_{u^0_L}$ for the
quark, $p_{u^{0*}_{L}}$ for the anti-quark, and $q$
for the intermediate squark. The factor $v (p_m, q, p_{n})$ is
given by the integration in the extra dimension of the respective
profiles and the propagator:
\beq v (p_{\tilde g }^n, q, p_{\tilde g }^m) &=& \mathcal{N}_{u}^2
(\mu_{f}, 0) \mathcal{N}_{\tilde g}(\mu_{g}, p_{\tilde g }^m)
\mathcal{N}_{\tilde g }(\mu_{g}, p_{\tilde g }^n ) \nonumber \\
&&\times \int_{z_{UV}
}^{z_{IR}} {dz} \int_{z_{UV} }^{z_{IR}} {d z'} ~ e^{(\mu_{g} -\mu_{f}) z}
z^{1/2 - c} h_L(p_{\tilde g }^m ,z)
z_{UV}^{-1}  \tilde{P}(q, z, z') \nonumber \\
&&\quad\quad\quad\quad\quad\quad\quad\quad\quad e^{(\mu_{g} -\mu_{f}) z'} z'^{1/2 - c} h_L(p_{\tilde g }^n,z')~, \eeq
here we use a rescaled scalar propagator, $ \tilde{P}(q, z, z')=
\left( {\frac{z}{z_{UV}}} \right)^{-3/2} \left(
{\frac{z'}{z_{UV}}} \right)^{-3/2} P(q, z, z')$, and we can write
the differential decay rate as:
\beq \frac{{d\Gamma_3 }}{{dE_u }} = g_5^4 ~ {v (p_{\tilde g }^m,
q, p_{\tilde g }^n)}
 {v (p_{\tilde g }^m, q, k_{\tilde g }^n)^\dagger}
~\frac{{E_u^2 (2E_u p_{\tilde g }^m - (p_{\tilde g }^m)^2 +
(p_{\tilde g }^n)^2 )^2}}{{32 p_{\tilde g }^m (2E_u  -
p_{\tilde g }^m)  \pi ^3 }} \frac{1}{8} {\rm Tr} [t^at^bt^bt^a] \label{total}\\
\nonumber
\eeq
where $E_{u_L^0 }$ is the energy of the (first) outgoing quark in
the initial gluino rest frame and $\frac{1}{8}{\rm Tr}
[t^at^bt^bt^a] = \frac{1}{8} C_A C_F^2 = 2/3$ for $SU(3)$. The
range of $E_{u_L^0 }$ is determined by the masses of the initial
and final gluinos:
\beq
 E_{u_L^0 \min }  &=& 0 ~,\label{E2min} \\
 E_{u_L^0 \max }  &=& \frac{1}{2} ~ p_{\tilde g }^m~
 \left( {1 - \frac{{(p_{\tilde g }^n )^2 }}{{(p_{\tilde g }^m)^2 }}}
 \right)~.
\label{E2max}\eeq
\begin{figure}[t]
\centering
\begin{tabular}{cc}
\includegraphics[scale=0.56]{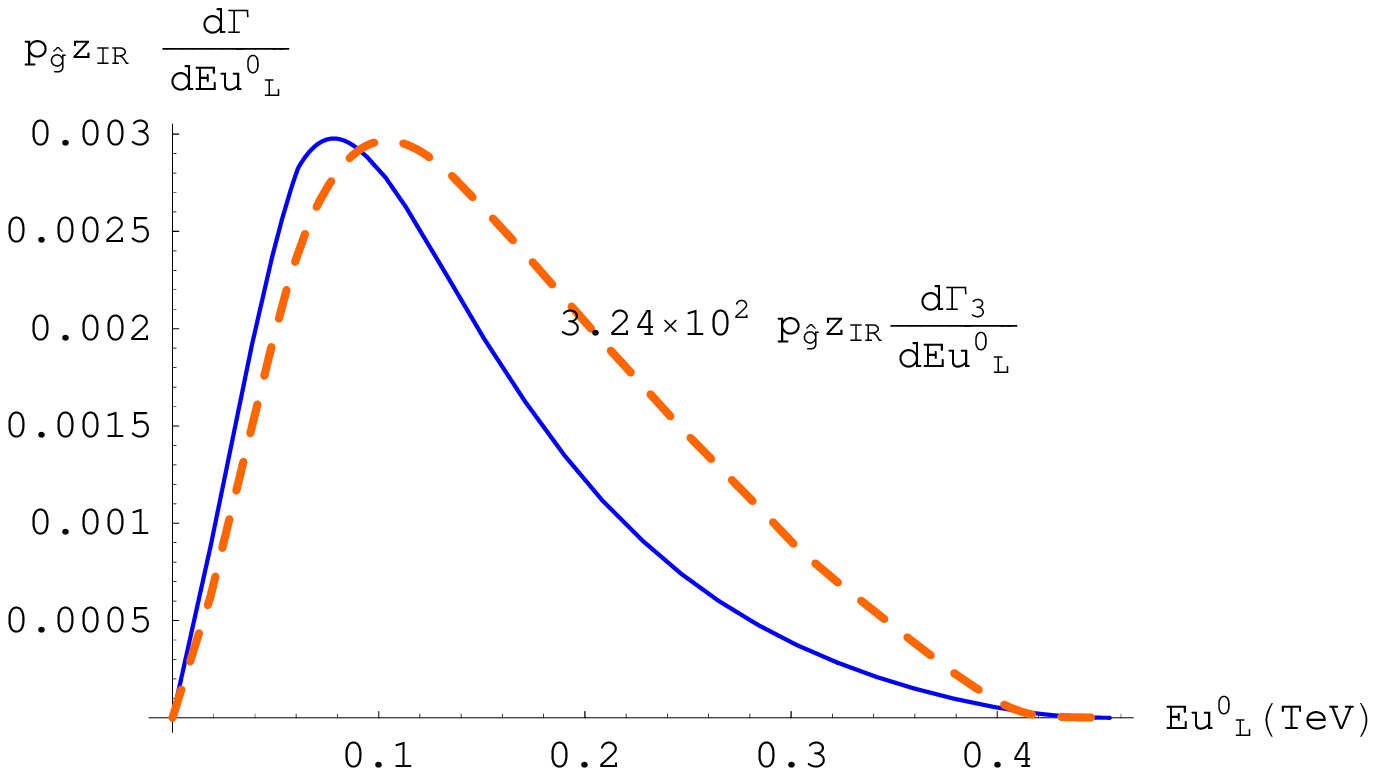}
&
\includegraphics[scale=0.56]{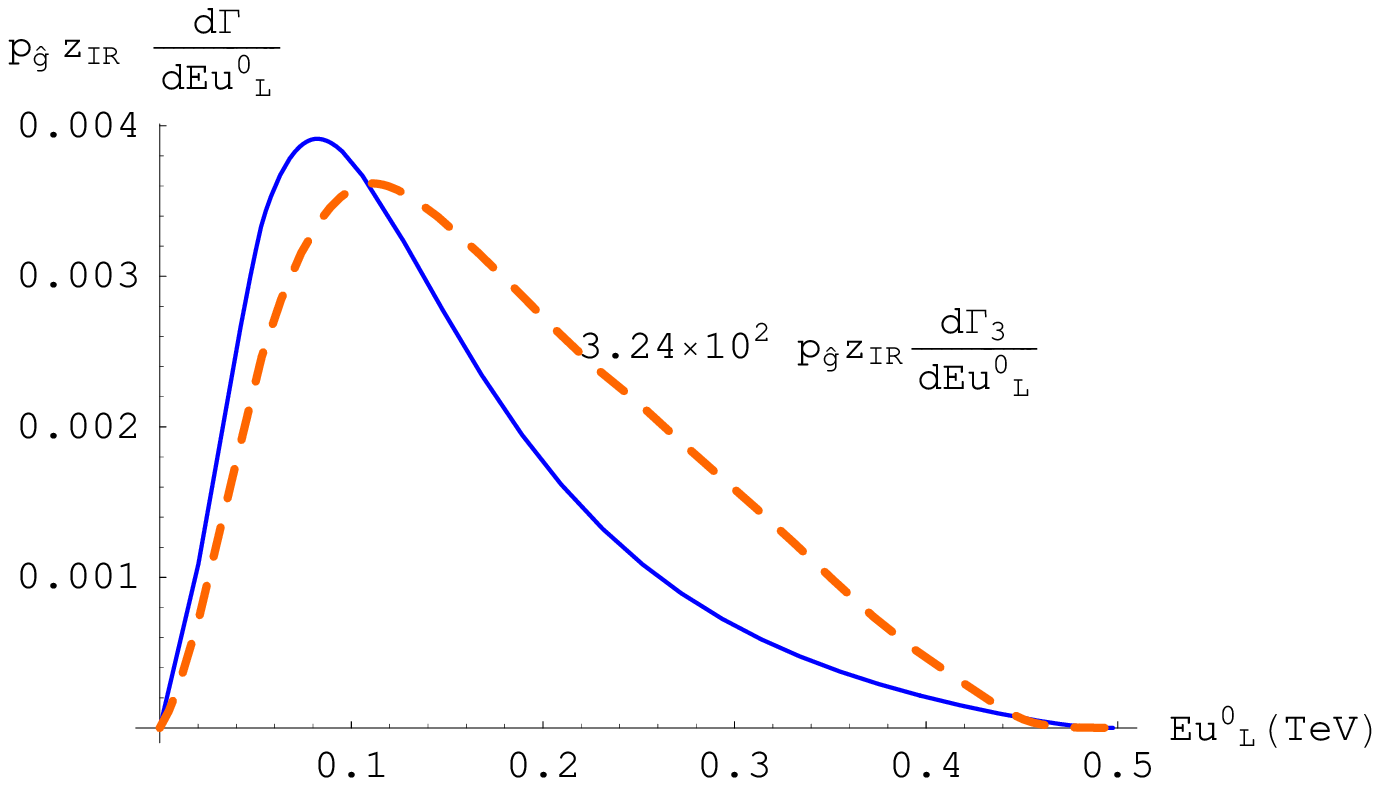}
\end{tabular}
\caption {Gluino decay rate with respect to the first quark
energy. The yellow dash line represents the virtual contribution
from three body  decay which is rescaled by a factor of $3.24
\times 10^2$ , while the blue line represents two body decay
result. In both plots, the initial KK gluino mass is $ p_{\tilde
g} =1.15 ~\mbox{TeV}$ and we use $g_5 = 0.108 ~\mbox{TeV}^{-1/2}$.
In the left one, we choose $z_{UV} = 10^{-3} ~ \mbox{TeV}^{-1}$,
$z_{IR} = 80 ~ \mbox{TeV}^{-1}$, $c= 0.5$, $\mu_{g}= 0.4
~\mbox{TeV}$ and $\mu_{f} = 0.5 ~\mbox{TeV} $; In the right one,
we choose $z_{UV} = 10^{-3} ~ \mbox{TeV}^{-1}$, $z_{IR} = 100 ~
\mbox{TeV}^{-1}$, $c=0.5$, $\mu_{g} = 0.3 ~\mbox{TeV}$ and
$\mu_{f} = 0.4 ~\mbox{TeV}$.}\label{twovsthree}
\end{figure}

The contribution to the differential decay rate as a function of
the outgoing quark energy from the virtual 3-body process
can be compared with the real 2-body contribution by summing
over all final KK gluinos and outgoing anti-quark energies
which are kinematically allowed. The result is shown in
Fig.~\ref{twovsthree}. The virtual 3-body process is the cut of a two-loop diagram
while the real 2-body process is the cut of a one-loop diagram, so
we expect that the 3-body decay rate should be suppressed with
respect to the 2-body decay rate by a 4D loop factor as long as the theory remains perturbative. The
shapes of the two contributions are also similar, but with the 3-body
contribution being slightly harder. Thus we can conclude that in this case, it is reasonable
to use the narrow-width approximation to calculate the energy
distributions of the visible particles coming from the
continuum decays by treating each decay step going to real states that subsequently
decay  independent of the details of the previous decay.

\section{Phenomenology of the Continuum Superpartner Decays}

\begin{figure}[t]

       \centering
       \begin{tabular}{cc}
       \includegraphics[scale=0.56]{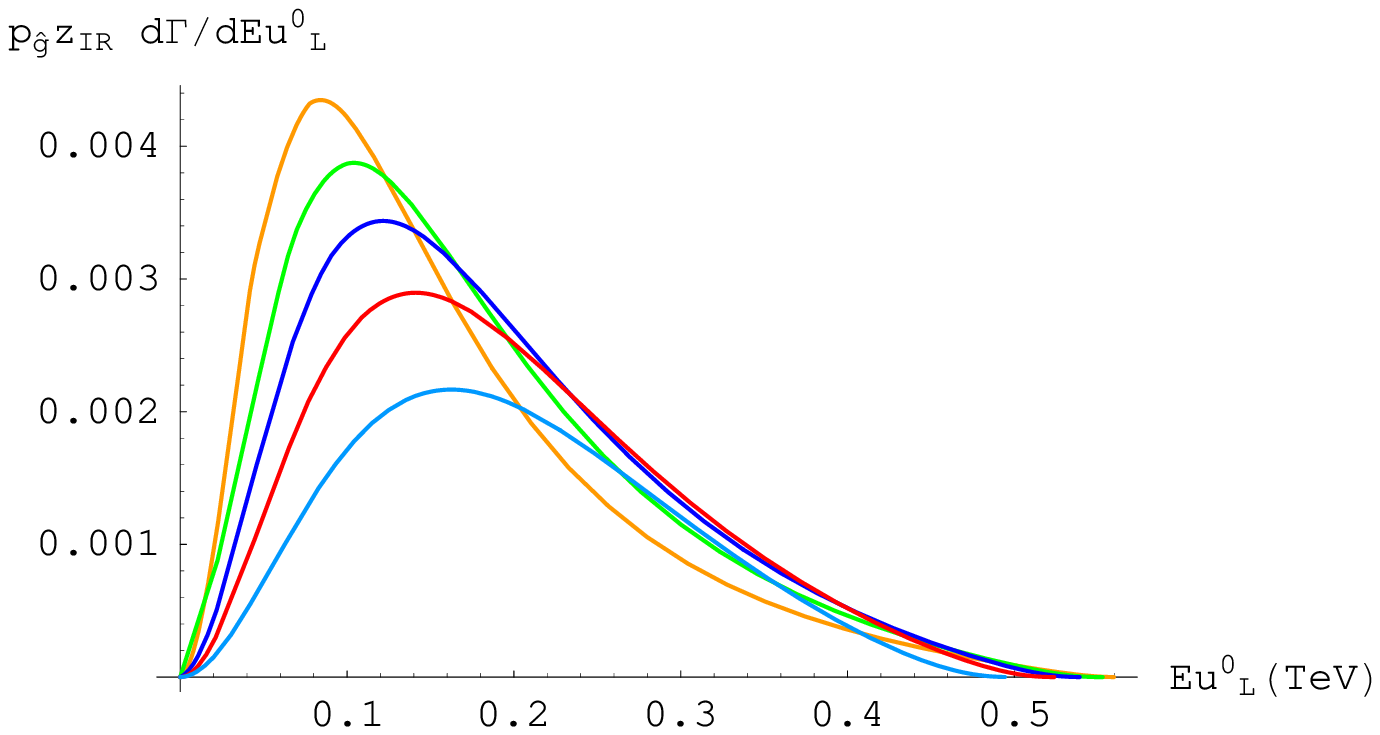}&
       \includegraphics[scale=0.56]{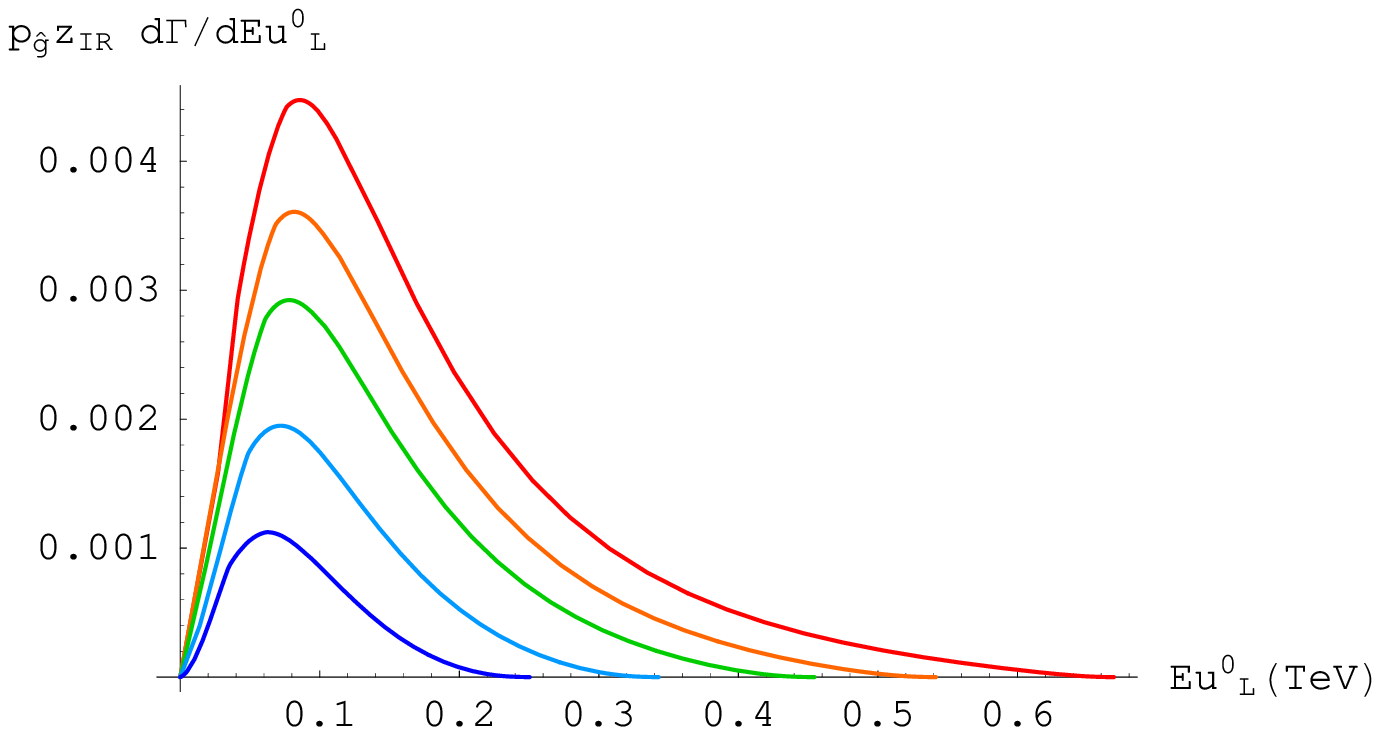}
        \end{tabular}

\caption{ 2-body gluino differential decay rate as a function of
the outgoing quark energy $ E_{u^0_{L}}$ in TeV. We have taken in
this example  $z_{UV} = 10^{-3} ~ \mbox{TeV}^{-1}$, $z_{IR} = 80 ~
\mbox{TeV}^{-1}$, $c= 1/2$ and $ g_5 = 0.108 ~ \mbox{TeV}^{-1/2}
$. In the figure on the left, we fixed the initial gluino KK-mass
at $ p_{\tilde g}  = 1.26 ~\mbox{TeV}$ and its mass-gap to
$\mu_{g} = 0.3 ~\mbox{TeV}$, and vary the squark mass gap by
$\mu_{f}= 0.40 ~, 0.43~, 0.46~, 0.50~, 0.56 ~\mbox{TeV} $. As can
be seen from the figure, the peak position decreases in magnitude
and shifts towards larger values of $E_{u^0_{L}}$. In the figure
on the right, we fixed $\mu_{g} = 0.4 ~\mbox{TeV}$ and $\mu_{f} =
0.5 ~\mbox{TeV} $, and vary the initial gluino KK-mass by $
p_{\tilde g} = 0.83 ~,0.97~ ,1.15~,1.29~, 1.52 ~ \mbox{TeV}$. In
this case the peak increases in magnitude with increasing $ p $
and its position remains roughly constant as a function of $
E_{u^0_{L}}$.}\label{width1}
\end{figure}
\begin{figure}[t]

       \centering
       \begin{tabular}{cc}
       \includegraphics[scale=0.56]{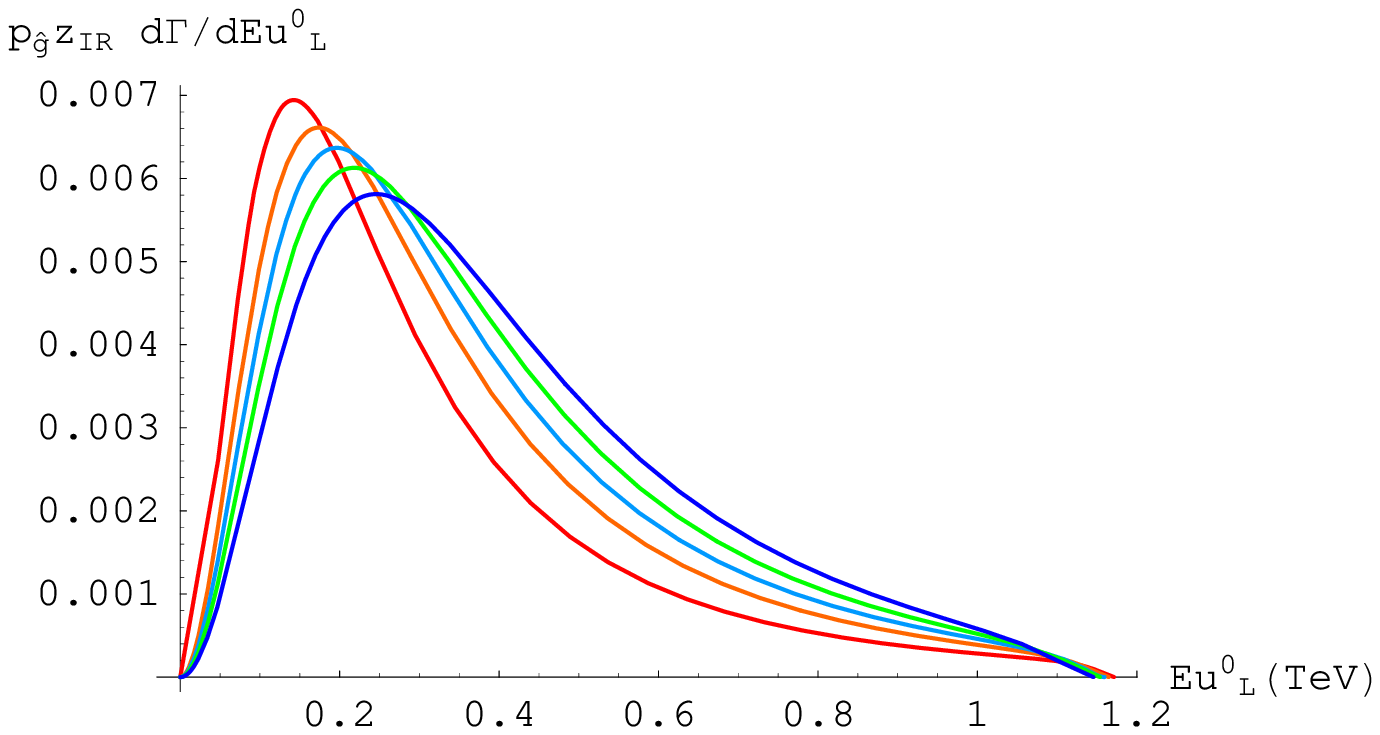}&
       \includegraphics[scale=0.56]{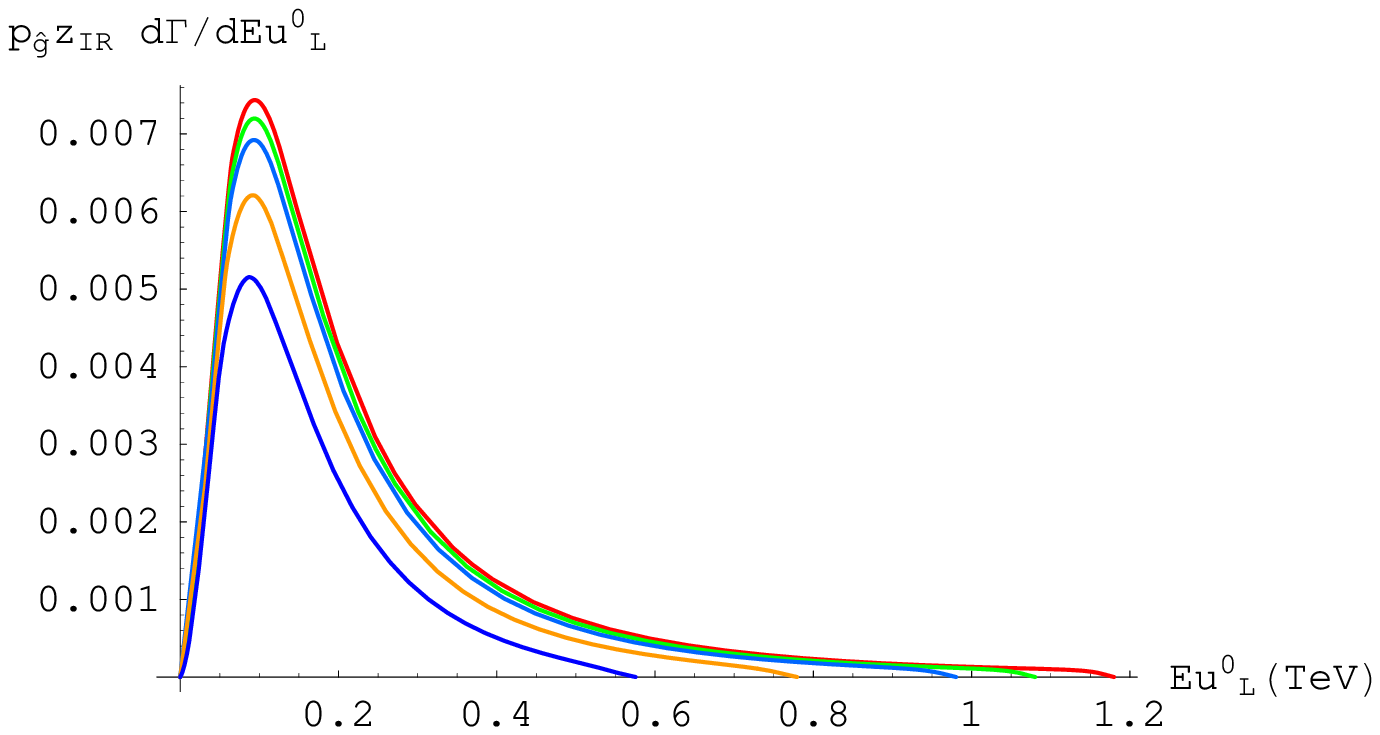}
        \end{tabular}
\caption{Same as Fig.~\ref{width1} for different sets of energies and mass gaps.
In the plot on the left, we fixed $ p_{\tilde g}= 2.4
~\mbox{TeV}$, and $\mu_{g} = 0.2 ~\mbox{TeV}$. We vary the squark
mass gap by $\mu_{f}= 0.36~, 0.40~, 0.43~,0.46~, 0.5 ~\mbox{TeV}
$. As can be seen, the peak decreases in magnitude and slightly
moves towards larger values of $ E_{u^0_{L}}$. In the figure on
the right, we fixed $\mu_{g} = 0.2 ~\mbox{TeV}$ and $\mu_{f} = 0.3
~\mbox{TeV} $, and vary the initial gluino KK-mass by $ p_{\tilde
g} = 1.24~, 1.62~, 2.01~ ,2.21~, 2.4 ~ \mbox{TeV}$. In this case,
the peak magnitude increases and its position remains roughly
constant with respect to  $ E_{u^0_{L}}$.}\label{width2}
\end{figure}
\begin{figure}[t]
\centering
\begin{tabular}{c}
\includegraphics[scale=0.6]{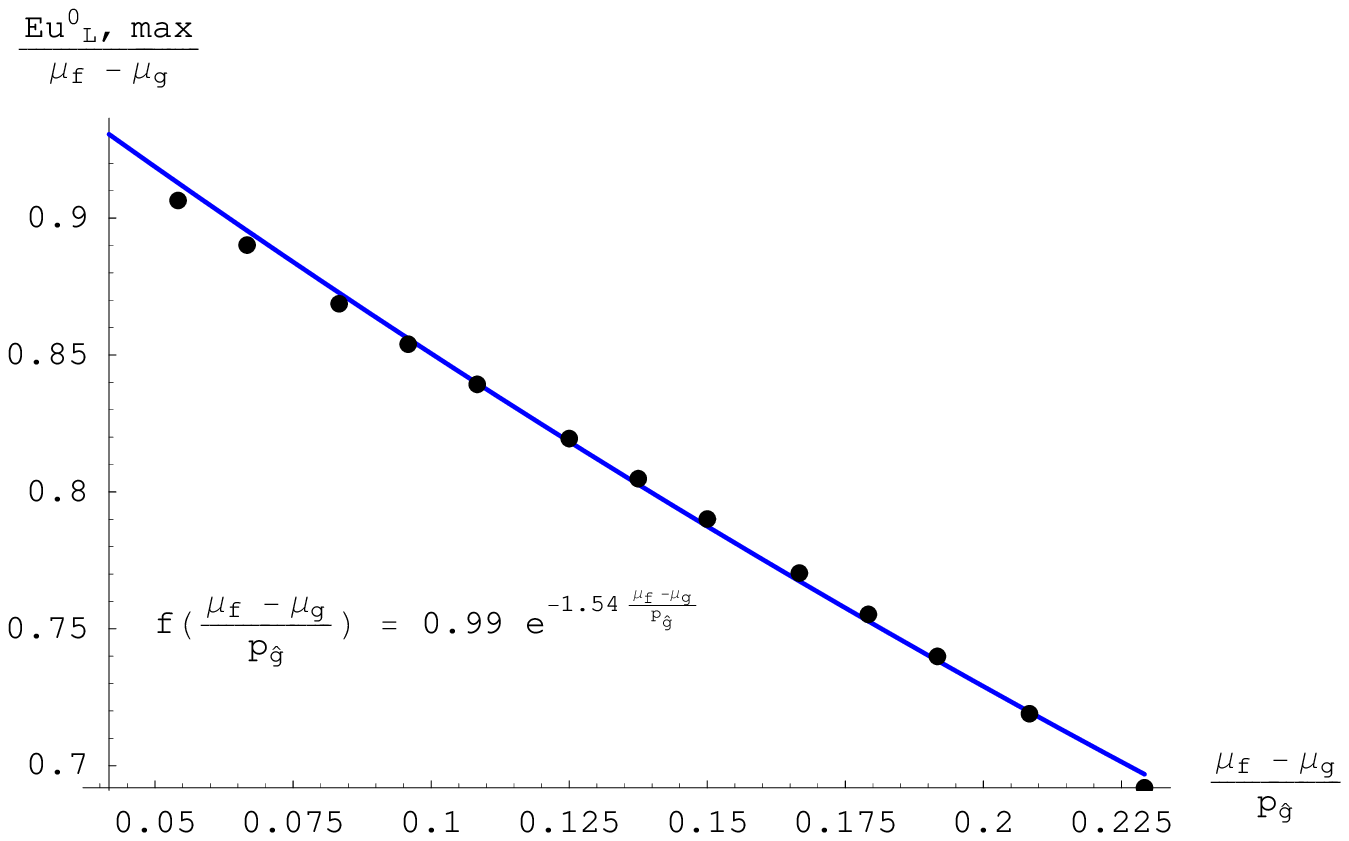} \\
\includegraphics[scale=0.6]{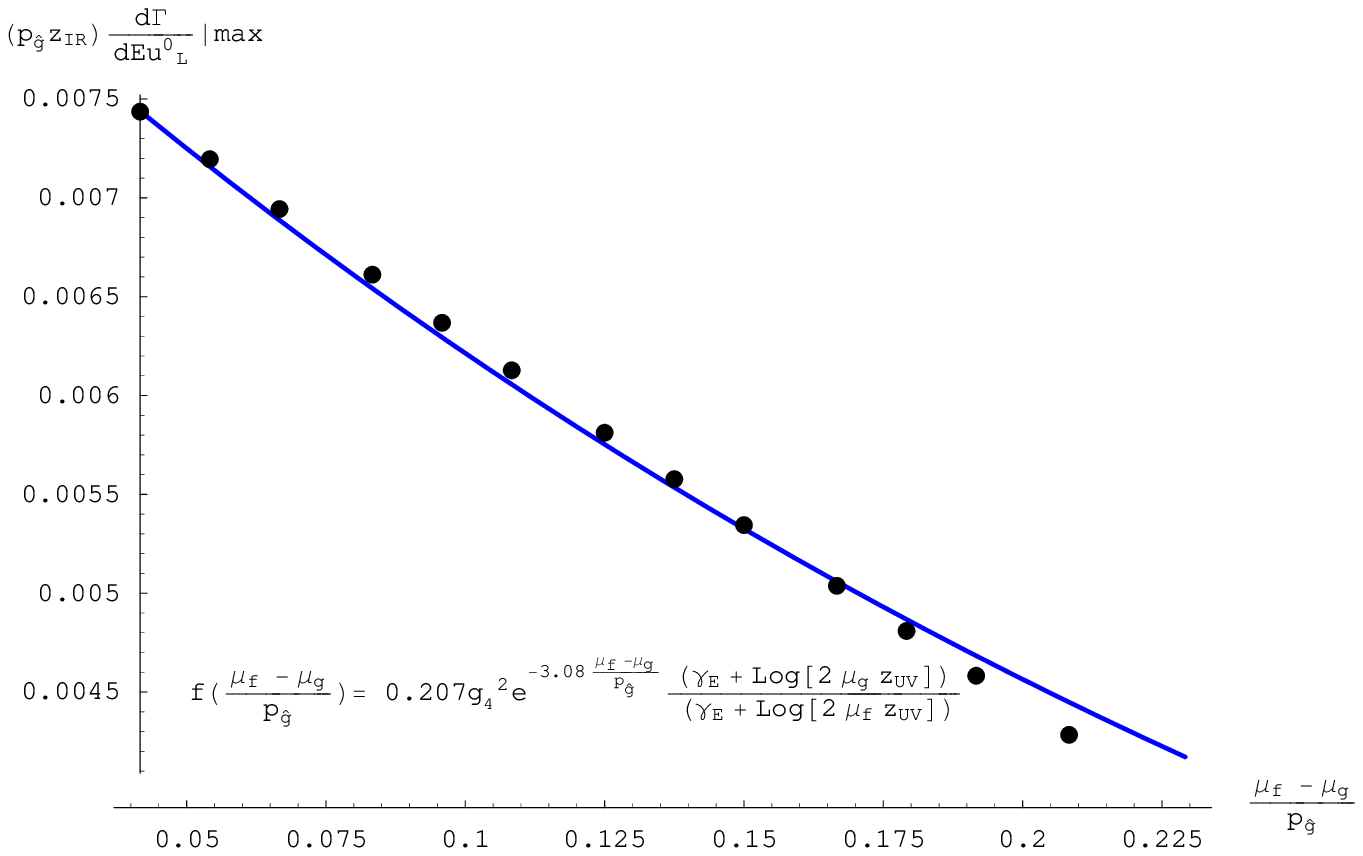}
\end{tabular}

\caption{$ \frac{E_{u^0_{L},max}}{\mu_{f}-\mu_{g}}$ vs.
$\frac{\mu_{f}-\mu_{g}}{p_{\tilde g}}$ and $(p_{\tilde
g}z_{IR})\frac{ d \Gamma}{dE_{u^0_{L}}}|_{max}$ vs.
$\frac{\mu_{f}-\mu_{g}}{p_{\tilde g}}$ evaluated at the peak
positions. We use $z_{UV} = 10^{-3} ~ \mbox{TeV}^{-1}$, $z_{IR} =
80 ~ \mbox{TeV}^{-1}$, $c= 0.5$ and $ g_5 = 0.108
~\mbox{TeV}^{-1/2} $ and fix $ p_{\tilde g}= 2.40 ~\mbox{TeV}$,
$\mu_{g} = 0.2 ~\mbox{TeV}$, while the squark mass gap is varied
from $\mu_{f}= 0.3 ~\mbox{TeV}$ to $\mu_{f} = 0.75 ~\mbox{TeV} $.
The plots display the fitted functions. }\label{peak}
\end{figure}
The comparison of the differential decay rate calculations for a
real 2-body process  and a virtual 3-body process in the previous
section, shows that it is reasonable to calculate the energy
distributions of the visible particles coming from the continuum
superpartner decays using the narrow-width approximation. This
greatly simplifies the phenomenological study of continuum
superpartners. As mentioned earlier, the continuum states tend to
decay to other continuum states that are nearby in invariant mass,
and so the ordinary particles tend to be emitted with soft
energies. This is the behavior that is expected in a CFT
\cite{Strassler,Polchinski,HofmanMaldacena,Csaki}. Thus if the
decay chain starts fairly high up in the continuum, then there is
usually a long decay chaining with an approximately spherical
distribution of energy \cite{Csaki}. In this section, we examine
how the energy distributions of the visible particles depend on
the parameters of the theory and the process.

As  the energy of the emitted quark $ E_{u^0_{L}}$ increases (or
what is the same, the squark mass $q_n$ decreases), the phase
space of the final particles increases while the vertex
$c(p_n,q_m)$ decreases. It is the competition between these two
factors that makes the differential decay rate have its maximum at
an quark energy around $ E_{u^0_{L},{\rm max}} \sim (\mu_{f}
-\mu_{g} )$ as can be seen in Figs.~\ref{width1} and \ref{width2}.
As a consequence, for a high-energy  initial state ($p_{\tilde g}
\gg \mu_f, \, \mu_g$), we expect order $p_{\tilde g}/(\mu_f
-\mu_g)$ particles coming out a decay chain. For fixed  $g_4$,
$c$, $z_{IR}$, and $z_{UV}$, there are only three parameters: the
initial gluino energy $p_{\tilde g}$ and the two mass gaps
$\mu_f$, $\mu_g$. Since there should be little dependence on
$z_{IR}$ and $z_{UV}$, as long as they are far away from the
energy scale interested, physical quantities can be expressed as
functions of the two dimensionless ratios of $p_{\tilde g}$,
$\mu_f$, and $\mu_g$ up to an overall normalization. In
Fig.~\ref{peak} we show the dependence of the peak position and
magnitude of the differential decay rate on the initial energy
$p_{\tilde g}$ and the mass gap difference $\mu_f -\mu_g$. From
these figures we can determine an approximate behavior for the
peak position, $E_{u^0_{L},{\rm max}}$ in the differential decay
rate as a function of $\mu_{f}-\mu_{g}$ and $p_{\tilde g}$. We
find that it can be parameterized by an exponential decaying
function with an overall normalization $A$ and a numerical
exponent $B$:
\beq E_{u^0_{L},{\rm max}}= A (\mu_{f}-\mu_{g}) ~ e^{- B\frac{(\mu_{f}-\mu_{g}
)}{ p_{\tilde{g}}}}\;.
\label{Emaxapprox}
\eeq
The numerical values of $A$ and $B$ are obtained by fitting
numerical data points, as shown in Fig.~\ref{peak}, where we find
that $A\approx 1$ and $B\approx 3/2$.  Using the plane wave
approximation for the wave functions of the incoming gluino and
outgoing squark, the coupling squared is proportional to $|c(p_m,
q_n) |^2 \propto \frac{1}{z_{IR}^2 ~(\mu_f -\mu_g)^2}$~ . Then one
can substitute Eq.(\ref{Emaxapprox}) into Eq.(\ref{twobody}),
using $\Delta E_{u^0_L}\approx \pi/z_{IR}$, to obtain an
approximate analytical expression  for
$(d\Gamma/dE_{u^0_{L}})|_{E_{u^0_{L},{\rm max}}}$ as a function of
$\mu_{f}$ and $\mu_{g}$:
\beq \left( p_{\tilde{g}}
z_{IR}\frac{d\Gamma}{dE_{u^0_{L}}}\right)\bigg|_{E_{u^0_{L},max}} = C
\cdot g_4^2 \frac{e^{- 2 B \frac{ (\mu_{f}-\mu_{g}
)}{p_{\tilde{g}}}}}{4 \pi ^2} \frac{( \log (1/(2 \mu_{g} z_{UV})
)-\gamma )}{( \log (1/(2 \mu_{f} z_{UV}) )-\gamma )}\nonumber\\
\label{dGdEmax}
\eeq
the overall normalization $C$ compensates for the inaccuracy of the plane wave
approximation, and we have written everything in terms of 4D quantities.
The dependence on $z_{UV}$
is only logarithmic and therefore mild. To give an idea of the validity of the approximation,
we plot against numerical data in the second plot of Fig.~\ref{peak}.
We see that (\ref{Emaxapprox}) and (\ref{dGdEmax}) provide reasonably good estimates
for the functional dependence.

\section{Conclusions}

Supersymmetry is one of the best motivated scenarios for new physics beyond the standard model at the TeV scale.
 For the past two decades it has been intensively searched for. Currently, the experiments at the LHC have placed very strong limits on the masses of the squarks and gluino to be above $\sim$ 1 TeV in the standard scenario~\cite{daCosta:2011qk,Khachatryan:2011tk,Chatrchyan:2011ek}. This means that either the superpartner spectrum is unnaturally heavy or the superpartners decay in unusual ways which escape the standard SUSY searches. As we showed in this paper, if the superpartners have continuous spectra, they tend to have long decay chains and produce many soft SM particles. This is a challenging scenario at the LHC because the soft particles may not pass the experimental cuts, and the signals could be buried in the QCD backgrounds. It may require a more specialized study to search for this kind of signal.

Na\"ively the phenomenological studies of continuum superpartners at colliders may appear to be formidable as the usual calculation techniques have only been developed for particles. Here we have seen that these methods can still be used in continuum calculations. In particular, we showed that the narrow-width approximation is still generally valid in the perturbative decay processes that we are interested in. This greatly simplifies the calculations because processes with long sequences of decays can be divided into individual steps involving real states and each step can be calculated independently. The easiest way to perform the calculations is to introduce a regularizing IR brane in the 5D picture which transforms the continuum into discrete KK modes, then one can carry out the calculations as in the usual particle case. As long as the KK modes have reasonably fine spacings, the results are basically independent of the position of the IR brane. In this way, continuum superpartners can also be implemented in the usual collider simulation tools in order for more detailed studies to develop new strategies to search for such kinds of exotic collider signals.

\section*{Acknowledgments}
We thank Jack Gunion and Markus Luty for useful discussions and comments. We also thank the Aspen Center for Physics and the Kavli Institute for Theoretical Physics where part of this work was completed. The authors were
supported by the US Department of Energy under contract
DE-FG02-91ER406746.

\appendix

\section{Equivalence of the Continuum and KK Limit}
\label{sec:appendix}

It is interesting to compare our expression for the propagator. Eq.~(\ref{pro}),
with the more familiar  KK-representation of the 5D propagator which can be expressed
as sum over the different KK-level propagators:
\beq  P_{KK}(q,z,z') = \sum\limits_n {\frac{\tilde f_L (m_n, z)~
\tilde f_L^{\dagger}(m_n, z')}{q^2 - m_n^2 - i ~ \epsilon }},
\label{KK}\eeq
where $\tilde f_L (m_n, z)$ is the normalized left-handed squark
wavefunction. This propagator corresponds to a real particle  when $q^2=m_n^2$ (aka going "on-shell") for
each particular KK-mode $n$. For momenta $q$ close to $m_n$, we
can use the approximation,
\beq
\frac{i}{q^2  - m_{n}^2  + i \epsilon} = \pi \delta (q^2  - m_{n}^2 ) +i\, {\mathcal P}(\frac{1}{q^2 -
m_{n}^2}) ,
\label{imag}
\eeq
where ${\mathcal P}$ denotes the Cauchy principal value. As the mass difference between
adjacent KK-levels $n$ and $m=n-1$ tends to zero (the continuum limit), $\Delta m_{n}=m_{n}-m_{m}\to 0$,
the sum over KK-levels in Eq.~(\ref{KK}) becomes an integral in the complex plane
over ``tightly squeezed" resonances which form a branch cut in the limit
$z_{IR}\rightarrow \infty $. Thus, we can identify the real (``on-shell")
contribution of the continuum propagator, Eq.~(\ref{pro}), by matching
the continuum limit of  Eq.~(\ref{KK}) with Eq.~(\ref{pro}). For that
purpose, we consider Eq.~(\ref{KK}) in the limit $\mu_{f}\ll q\ll1/z_{UV}$.
Let us take $z=z'=z_{UV}$ and using Eq.~(\ref{fL}) and Eq.~(\ref{NL}), in the case $c=1/2$ for simplicity, we find that:
\beq \left| {\tilde f_L(m_n,z_{UV} )} \right|^2  &\to& \left|
{\mathcal{N}_L \left( { - \frac{{M(0,0,2i m_n z_{UV} )}}{{W(0,0,2i
m_n z_{UV} )}}W(0,1,2i m_n z_{UV} )} \right)} \right|^2  \nonumber \\
&\to&  \frac{\pi}{ z_{IR}} \frac{1}{\sqrt{m_n^2} z_{UV}
\left(\left(\log \left(\frac{m_n}{2}\right)+\log (z_{UV})+\gamma_E
\right)^2+\frac{\pi^2}{4}\right)} \label{fLB}. \eeq
By identifying $\Delta m_n = \pi / z_{IR} $ and using Eq.~(\ref{imag}),
we find that:
\beq && \mbox{Im}[P_{KK}(q,z_{UV},z_{UV})]\to\sum\limits_n \left|\tilde f_L (m_n, z_{UV})\right|^2 ~
\pi\delta (q^2 - m_{n}^2 ) \nonumber  \\
&\to& \frac{\pi /2 }{q^2 z_{UV} \left(\left(\log
\left(\frac{q}{2}\right)+\log (z_{UV})+\gamma_E
\right)^2+\frac{\pi^2}{4}\right)}=
\mbox{Im}[\frac{\Sigma_{F_c}}{z_{UV}}]. \eeq
One can also see the equivalence between the KK picture in the limit $z_{IR}\to +\infty$ and
the continuum picture by comparing the differential decay rate for large $z_{IR}$ in the KK picture versus the
continuum case where one still keeps
the particle behavior for the initial gluino state, but replaces the
final state squark phase space, by an unparticle phase space~\cite{Georgi}.

For that purpose, let us calculate the continuum squark phase
space for $c\neq 1/2$ and $c=1/2$. In order to get the phase space
of the squark final state, we calculate first the spectral
function $\rho(p^2)=2 \cdot {\rm Im}(i\Delta(p^2))$, where
$\Delta(p^2)$ is the squark correlator whose expression was
calculated in Ref.~\cite{Continuum}. As was done in previous
calculations, we are interested in the case when the momenta
involved are much bigger than the squark's mass gap, $p\gg
\mu_{f}$. Using the expressions of the correlator in the limit
when $p z_{UV}\ll 1$ we then find that for $-1/2<c < 1/2$,
\begin{equation}
\rho(p^2)=\frac{2 B(p^2)}{A(p^2)^2+B(p^2)^2},
\end{equation}
where
\begin{eqnarray}
A(p^2)&=&
\frac{p^2\epsilon^{1-2c}}{1-2c}-\frac{2^{-1+2c}p^2(p^2-\mu_{f}^2)^{-1/2+c}\cos(\pi(-\frac{1}{2}+c))\Gamma(1-2c)\Gamma(c)}
{\Gamma(2c)\Gamma(1-c)}, \nonumber \\
B(p^2)&=&
-\frac{2^{-1+2c}p^2(p^2-\mu_{f}^2)^{-1/2+c}\sin(\pi(-\frac{1}{2}+c))\Gamma(1-2c)\Gamma(c)}
{\Gamma(2c)\Gamma(1-c)},
\end{eqnarray}
and that for $c=1/2$,
\begin{equation}
\rho(p^2)=\frac{2 \cdot \frac{\pi}{2}
}{(\gamma_E+\log(\frac{\epsilon}{2}\sqrt{p^2-\mu_{f}^2}))^2
p^2+\frac{\pi^2}{4}p^2}.
\end{equation}
The continuum squark phase
space is then given by,
\begin{equation}
d\Phi_{\tilde{e}}=\theta(p^{0})\theta(p^2-\mu_{f}^2)\rho(p^2),
\end{equation}
where $\theta(p)$ is the Heaviside step function.
On the other hand, the phase space for the outgoing quark is just the usual particle one,
\begin{equation}
d\Phi_e=2\pi \theta(p^{0})\delta(p^2).
\end{equation}
With all this information, we are ready to calculate the 2-body
phase space integral.~\footnote{Even though the vertex depends on
the momenta of the particles, it only ends up depending on the
quark energy $E_u$ and the initial gluino energy
$p_{\tilde{g}}^0$.} Let us name the momentum of the incoming
gluino as $p_{\tilde{g}}$, and of the outgoing quark and
squark momenta as $p_{u}$ and $p_{\tilde{u}}$. Then the 2-body
phase space integral is given by
\begin{equation}
\Pi_2=\int(2\pi)^4\delta^{(4)}(p_{\tilde{g}}-p_{\tilde{u}}-p_u)2\pi\delta(p_u^2)\theta(p_u^0)\rho(p_{\tilde{u}}^2)
\theta(p_{\tilde{u}}^0)\theta(p_{\tilde{u}}^2-\mu_{f}^2)\frac{d^4p_u}{(2\pi)^4}\frac{d^4p_{\tilde{u}}}{(2\pi)^4}.
\end{equation}
We can perform the integral over the part of the quark phase space
using that,
\begin{equation}
\int 2\pi\delta(p_u^2)\theta(p_u^0)\frac{d^4p_u}{(2\pi)^4}=\int
\frac{d^3p_u}{(2\pi)^3 2E_u}\bigg|_{p_u^0>0}.
\end{equation}
Using the 4-dimensional delta function to perform the integrals over
$d^4p_{\tilde{u}}$ and in the center of mass frame (CM) of the
gluino, we can trivially perform the angular integration,
$d^3p_u=E_u^2dE_ud\Omega$, so that we obtain,
\begin{equation}
\Pi_2=\frac{1}{2\pi^2}\frac{E_u^2}{2E_u}\rho((p_{\tilde{g}}^0)^2-2p_{\tilde{g}}^0E_u)
\theta((p_{\tilde{g}}^0)^2-2p_{\tilde{g}}^0E_u-\mu_{f}^2)dE_u .
\end{equation}
Once we have the expression for the phase space integral, we can
write the differential decay rate as,
\begin{equation}
d\Gamma=\frac{|\mathcal{M}|^2}{2p_{\tilde{g}}^0}\Pi_2 .
\end{equation}
\begin{figure}[t]
\centering
\begin{tabular}{c}
\includegraphics[scale=0.6]{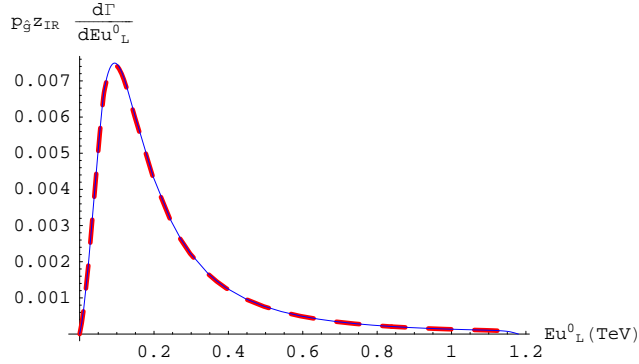}
\end{tabular}

\caption{ We use $z_{UV} = 10^{-3} ~ \mbox{TeV}^{-1}$, $z_{IR} =
80 ~ \mbox{TeV}^{-1}$, $c= 0.5$ and $ g_5 = 0.108
\mbox{TeV}^{-1/2} $ and fix $ p_{\tilde g} = 2.40 ~\mbox{TeV}$,
$\mu_{g} = 0.2 ~\mbox{TeV}$, and  $\mu_{f}= 0.3 ~\mbox{TeV}$. The
plots compare the continuum method (red dashed curve) with the KK mode
method (blue curve), they are almost identical.}\label{compare}
\end{figure}
Averaging over the initial and summing over the final spin and
color, we find that \beq |\mathcal{M}|^2=2 |\tilde
c(p_{\tilde{g}}^0,E_u,c)|^2 p_{\tilde{g}}^0 E_u{\rm
tr}[t^at^a]/8~, \eeq where \beq \tilde c(p_{\tilde{g}}^0,E_u,c) =
\frac{1}{z^{1/2}_{UV}}\cdot\frac{c(p_{\tilde{g}}^0,E_u,c)}{\tilde
f(q, z_{UV})}~, \eeq is the 4D effective vertex obtained from
integrating over the profiles in 5D, divided by the normalized
squark profile evaluated at the UV brane, and multiplied by
$\frac{1}{z^{1/2}_{UV}}$ to be dimensionless, with
$c(p_{\tilde{g}}^0,E_u,c)$ shown in Eq.~(\ref{coupling}). Thus we
arrive at the formula,
\begin{equation}
\frac{d\Gamma_2}{dE_u}=\frac{E_u^2}{4\pi^2}|\tilde
c(p_{\tilde{g}}^0,E_u,c)|^2
\rho((p_{\tilde{g}}^0)^2-2p_{\tilde{g}}^0E_u)\theta((p_{\tilde{g}}^0)^2-2p_{\tilde{g}}^0E_u-\mu_{f}^2).
\label{dGdE}
\end{equation}

In Fig \ref{compare} we plot the KK differential
decay rate, Eq.~(\ref{twobody}), and the continuum decay rate,
Eq.~(\ref{dGdE}), as a function of the final quark energy $E_u$. They are in excellent agreement for large enough $z_{IR}$.

\end{document}